\documentclass[aps,prb,twocolumn,showpacs,groupedaddress]{revtex4}
\usepackage{graphicx}
\usepackage{amsmath,amssymb}

\newcommand{\be}{\begin{equation}}
\newcommand{\ee}{\end{equation}}
\newcommand{\bea}{\begin{eqnarray}}
\newcommand{\eea}{\end{eqnarray}}
\newcommand{\ba}{\begin{eqnarray*}}
\newcommand{\ea}{\end{eqnarray*}}

\newcommand{\dis}{\displaystyle}

\newcommand{\fract}[2]{\frac{\dis #1}{\dis #2}}

\newcommand{\eqn}[1]{(\ref{#1})}

\newcommand{\bw}{\begin{widetext}}
\newcommand{\ew}{\end{widetext}}

\def\v2o3{V$_2$O$_3$}

\begin{document}

\title{Low-temperature magnetic ordering and structural distortions in
  Vanadium~Sesquioxide (V$_2$O$_3$)}

\author{Daniel Grieger}
\affiliation{SISSA, Via Bonomea 265, 34136 Trieste, Italy}
\author{Michele Fabrizio}
\affiliation{SISSA, Via Bonomea 265, 34136 Trieste, Italy}

\begin{abstract}
  Vanadium Sesquioxide (V$_2$O$_3$) is an antiferromagnetic insulator
  below $T_{\mathrm N}\approx$~155~K. The magnetic order is not of C-
  or G-type as one would infer from the bipartite character of the
  hexagonal basal plane in the high-temperature corundum structure. In
  fact, the N\'eel transition is accompanied by a monoclinic
  distortion that makes one bond of the honeycomb plane inequivalent
  from the other two, thus justifying a magnetic structure with one
  ferromagnetic bond and two antiferromagnetic ones.  We show here
  that the magnetic ordering, the accompanying monoclinic structural
  distortion, the magnetic anisotropy and also the recently discovered high-pressure monoclinic
  phase, can all be
  accurately described by conventional electronic structure
  calculations within GGA and GGA+U. Our results are in line with DMFT
  calculations for the paramagnetic phase~\cite{pot07}, which predict that 
  the insulating character is driven by a correlation-enhanced crystal
  field splitting between $e^\pi_g$ and $a_{1g}$ orbitals that 
  pushes the latter above the chemical potential.  We find that the
  $a_{1g}$ orbital, although almost empty in the insulating phase, is
  actually responsible for the unusual magnetic order as it leads to
  magnetic frustration whose effect is similar to a next-nearest-neighbor exchange in a 
  Heisenberg model on a honeycomb lattice.

\end{abstract}


\maketitle

\section{Introduction}

For more than forty years, the phase diagram of
Chromium/Titanium-doped Vanadium Sesquioxide~\cite{mcw71,mcw73}
(V$_2$O$_3$) has gathered great interest, especially because of its
isostructural high-temperature paramagnetic metal to paramagnetic
insulator transition, which is by now considered the prototypical
realization of a genuine Mott transition, i.e. not corrupted by any
symmetry breaking. Relatively less attention has instead been paid on
the low temperature antiferromagnetic phase of V$_2$O$_3$. Indeed,
within a certain doping/pressure range, Vanadium Sesquioxide undergoes
a magnetic phase transition~\cite{foe46} below a critical N\'eel
temperature $T_{\text{N}}$,~\cite{foe46,mcw70,moo70} which is around
155 K for undoped V$_2$O$_3$.~\cite{der70_2} Since magnetism in a
strongly-correlated material is just a side effect of Mott's
localization, scientific interest in V$_2$O$_3$ has mostly focused so
far on the latter phenomenon rather than on the mechanism
that produces the experimentally
observed magnetic order. Indeed, the magnetic structure in V$_2$O$_3$ rises a number
of intriguing questions most of which are still awaiting an answer.

\begin{figure}[ht]
\centering
\includegraphics[width=5cm,clip]{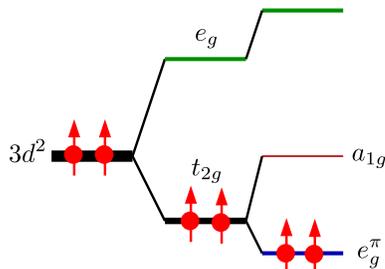}
\vspace{-0.5cm}
\caption{\label{levels} (Color online) $d$ levels and their occupancy of a hypothetical isolated Vanadium atom in the trigonal field of the high-temperature corundum structure.}
\end{figure}

In V$_2$O$_3$, each Vanadium atom has two electrons within the $t_{2g}$ orbitals of the cubic-split $d$-shell, 
as schematically shown in Fig.~\ref{levels}. In the high-temperature corundum structure, the trigonal field further splits the $t_{2g}$ into a lower $e^\pi_g$ doublet and a higher $a_{1g}$ singlet. In the extreme Mott localized scenario, the two electrons 
would occupy the $e^\pi_g$ orbitals and be coupled into a spin $S=1$ configuration in accordance with Hund's rules, see Fig.~\ref{levels}.
This idealized picture, each $e^\pi_g$ singly occupied and the $a_{1g}$ empty, is not far from what most recent LDA+DMFT calculations predict.\cite{pot07,gri12} If we assume legitimate to discard the $a_{1g}$ contribution to the low-energy processes that control the coupling between the $S=1$ localized moments, 
we must conclude that the virtual hopping of $e^\pi_g$ electrons between nearest neighbor sites gives rise to 
a conventional antiferromagnetic super-exchange within the honeycomb basal plane, whose bipartite character would then lead to a N\'eel two-sublattice antiferromagnetism. Moreover, since the $e^\pi_g$ 
orbitals are non-bonding along the $c$-axis perpendicular to the hexagonal plane, we would expect an 
antiferromagnetic order either of the G- or the C-type, which we shall hereafter denote as "simple"~\cite{ezh99} and "layered" antiferromagnetic structures, respectively. 

In reality, the experimentally observed magnetic structure~\cite{moo70}, which we shall refer to as 
"true", is completely different. Along the $c$-axis, the two nearest neighbor Vanadium atoms are coupled ferromagnetically, not in disagreement with the above expectation. In contrast, among the three bonds connecting one Vanadium atom to its nearest neighbors within the honeycomb basal plane, only two are antiferromagnetic but the remaining one is ferromagnetic. This  
phase is accompanied by a monoclinic distortion which goes along with the magnetic structure, 
making one hexagonal bond inequivalent from the other two. However, slightly contradicting results
about the exact influence onto the respective bond-lengths can be
found in literature, from shortening the antiferromagnetic
bonds,~\cite{der70_2,ten03} as one would reasonably expect, to the opposite.~\cite{roz02}

The natural issue that arises is why this complicated "true" magnetic order should be 
energetically favorable with respect to the "simple" or "layered" structures in view of 
the additional energy cost of the monoclinic distortion. Up to now, this simple question has still
not found a satisfactory answer.

\begin{figure}[ht]
\centering
\includegraphics[width=8.5cm]{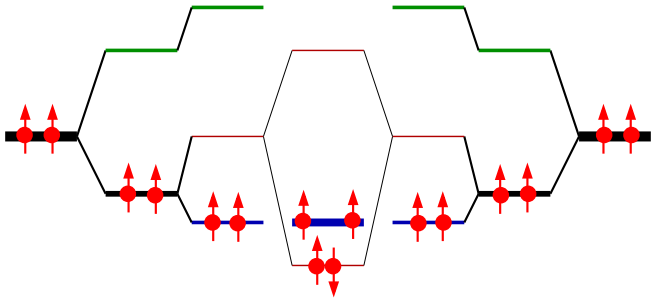}
\caption{\label{CNRlevels} (Color online) The "dimer" building block of Castellani, Natoli and Ranninger, with its electronic configuration.}
\end{figure}
The first attempt to explain the observed magnetic structure was performed 
in a series of papers by Castellani, Natoli and Ranninger (CN\&R).~\cite{cas78,cas78_2,cas78_3} Their starting point 
was not the atomic limit of Fig.~\ref{levels}, but the molecule of two nearest neighbor Vanadium atoms along the 
$c$-axis, which we shall refer to as a "dimer". The $a_{1g}$ orbitals form a covalent bond along the $c$-axis that falls below the $e^\pi_g$ levels, see Fig.~\ref{CNRlevels}. The lowest electronic configuration 
consists then of two electrons in a spin-singlet configuration occupying the $\sigma$-bond, and 
the remaining two coupled into a spin-triplet configuration within the $e^\pi_g$ levels. The residual 
fourfold orbital degeneracy besides the threefold spin degeneracy was exploited to build a spin-orbital 
Kugel'-Komskii~\cite{kug82} type of Heisenberg model, whose mean-field solution in a certain parameter range reproduces the observed magnetic structure and simultaneously predicts an orbital ordering. 
The CN\&R's scenario implies that each Vanadium has spin $S=1/2$, while the dimer has $S=1$. In order to explain the observed magnetic moment larger than one Bohr magneton, CN\&R invoked an exchange polarization of the $a_{1g}$ electrons. 

The "dimer" building block was later questioned on the basis of x-ray absorption measurements
~\cite{par00} and of {\sl ab-initio} LDA+U calculations,~\cite{ezh99} both supporting a scenario in which 
each Vanadium is in a spin $S=1$ configuration rather than spin-1/2. This conclusion was further reinforced by x-ray resonant elastic scattering measurements showing that the observed moment of 1.2 $\mu_B$ has both a spin contribution $2\langle S\rangle \simeq 1.7$ as well as an orbital one 
$\langle L \rangle \simeq -0.5$.~\cite{pao99} All these novel results stimulated attempts to reexamine   
the CN\&R model in terms of $S=2$ dimers (each Vanadium in a spin-triplet state) rather than $S=1$ as 
in the original formulation, see e.g. Refs. \onlinecite{mil00}, \onlinecite{mat02} 
and \onlinecite{tan02}, although this list is by no means exhaustive. 

At the meantime, the same belief that the V$_2$ dimer is the relevant building block to explain the magnetic structure started to be questioned,\cite{elf03,sah09,mat09} until most recent LDA+DMFT 
calculations~\cite{pot07,gri12} have finally come back to the atomic scenario of Fig. \ref{levels} as the most plausible one for the insulating phases of V$_2$O$_3$. Nevertheless, important issues remain 
open, which might escape from very accurate but still approximate techniques like LDA+U or 
LDA+DMFT.~\cite{mat05,mat09} 

An important one is the aforementioned sizable orbital contribution to the magnetic moment.~\cite{pao99} We observe that the $t_{2g}$ orbitals can make available at most an orbital moment $L=1$. Therefore the observed $\left|\langle L\rangle\right|\sim 0.5$ 
is a substantial part of it, which cannot be justified within the atomic limit of Fig.~\ref{levels}, since 
the $e^\pi_g$ alone are not spin-orbit active, but can be explained within the dimer scenario.~\cite{tan02,mat05} 

Equally intriguing remains the monoclinic distortion accompanying the magnetic order. 
As we mentioned, if we assume the atomic limit of Fig.~\ref{levels} and neglect contributions from 
$a_{1g}$ orbitals, only "simple" or "layered" magnetic structures can be stabilized. 
More realistic LDA+U calculations by Ezhov {\sl et al.}~\cite{ezh99} show that in the corundum structure the lowest energy magnetic configuration is indeed the "simple" one, lower by 5 K than the "true" structure. Since the monoclinic distortion costs elastic energy, it is not easy to conceive on the basis of these calculations why V$_2$O$_3$ would distort to stabilize a phase that in the undistorted crystal is higher in energy. Tight-binding Hartree-Fock calculations by Perkins {\sl et al.}~\cite{mat09} performed with hopping matrix elements of symmetry appropriate to the corundum phase find that the "true" structure can have lower energy than the "simple" one even in the undistorted lattice, though in a quite narrow 
region of parameter space. Similar conclusions are obtained within the dimer model.\cite{mil00,mat02} 

In summary, in spite of many attempts performed in the last forty years, the real cause of the observed magnetic order and concomitant monoclinic distortion in V$_2$O$_3$ is still elusive and we believe it is 
worth trying to shed further light, which we hope to do in the present study. The approach we shall adopt 
is mainly plain density
functional theory (DFT) and its DFT+U extension to strong electronic
correlations, which is especially suitable for antiferromagnetic
insulators. In this sense, this work is partially an
extension of the pioneering one by
Ezhov {\sl et~al.}\cite{ezh99} Furthermore, model studies are shown to
support the findings from these realistic theories.

The paper is organized as follows: Section~\ref{Sec:DFT} summarizes
the density functional theory (DFT) picture of V$_2$O$_3$ regarding
magnetism and structural distortions. Section~\ref{Sec:DFT+U} enhances
this picture by strong electronic correlations as described by DFT+U
approaches. Section~\ref{Sec:Orient} focuses on the orientation of the
magnetic moments as described by DFT+U as well as by an analytic
picture. Section~\ref{Sec:Models} finally aims at finding reasons for
the type of magnetic ordering realized in V$_2$O$_3$ by considering
suitable model studies.

\section{DFT studies\label{Sec:DFT}}

It is obvious that electronic correlations do play a crucial role in 
the physics of V$_2$O$_3$, as demonstrated in detail by several earlier 
studies,~\cite{ezh99,hel01,pot07,gri12} to which the
antiferromagnetic ground state makes no exception. However, already
density functional theory (DFT) in its generalized-gradient
approximation (GGA) to the energy functional (here in its PBE
parametrization~\cite{per96}), in spite of being known to fail for
several strongly correlated systems, can give some important insights
into this state. It is understood that one cannot expect an accurate
description of all the observed properties by plain GGA, but it will
be shown to be a useful starting point for all further considerations.

The following DFT calculations have been performed with the {\sc
  Quantum ESPRESSO} code~\cite{gia09} using ultrasoft pseudopotentials
(V.pbe-n-van.UPF and O.pbe-van\_ak.UPF from
http://www.quantum-espresso.org). To account for the monoclinic
distortion and magnetic ordering, a supercell containing 8
Vanadium atoms is used. Its symmetry properties are described by the
monoclinic space group $I2/a$,~\cite{der70,der70_2} whose lattice
vectors $(\mathbf{a}_{m}, \mathbf{b}_{m}, \mathbf{c}_{m})$ can be built from the original high-temperature
corundum structure lattice vectors $(\mathbf{a}_{\mathrm H},
\mathbf{b}_{\mathrm H}, \mathbf{c}_{\mathrm H})$ in hexagonal
notation as follows:
\begin{equation}
\left (
\begin{array}{c}
\mathbf{a}_{m}\\
\mathbf{b}_{m}\\
\mathbf{c}_{m}
\end{array}
\right ) = \left (
\begin{array}{ccc}
\frac 23 & \frac 43 & \frac 13\\
1 & 0 & 0\\
\frac 13 & \frac 23 & - \frac 13
\end{array}
\right ) \left (
\begin{array}{c}
\mathbf{a}_{\mathrm H}\\
\mathbf{b}_{\mathrm H}\\
\mathbf{c}_{\mathrm H}
\end{array}
\right ).
\end{equation}
If not stated otherwise, the length of the unit vectors will not be
altered throughout the following calculations, but retained at its
experimentally reported value at ambient conditions\cite{der70}.

\subsection{The paramagnetic solution}

The most basic GGA setup that can be built for V$_2$O$_3$ is a calculation
with enforced paramagnetism. If done in the enlarged 8-site unit cell
and allowing for relaxation of the atomic positions, its solution has
the noteworthy peculiarity that it incorporates a monoclinic
distortion.  
In spite of the elastic energy cost that is associated to any kind of
lattice distortion, the energy gain compared to the relaxed paramagnetic corundum structure is as large as 25~meV per Vanadium atom. The distortion is characterised by two nearest-neighbor bonds
being lengthened  and one shortened in the hexagonal $a_\text{H}b_\text{H}$~plane. 
Anticipating results of the next section, we mention that the same kind of distortion occurs in the 
antiferromagnetic metal phase. However, if GGA is supplemented by a Hubbard $U$, above a 
threshold value an antiferromagnetic insulating phase is established, which still has monoclinic distortion  
but with two bonds shortened and one lengthened, hence opposite to that in the metal phases, either paramagnetic or antiferromagnetic.

The large energy gain corresponds to a substantial lattice distortion,
with the length of the shortened bond in the
$a_\text{H}b_\text{H}$~plane of $2.56$~\AA{} and thus even slightly
smaller than that of the Vanadium dimer along the
$c_\text{H}$~direction of $2.63$~\AA. Comparing with the length of the
enlarged bonds in the $a_\text{H}b_\text{H}$~plane of about $3.0$~\AA,
one notices that the structure becomes similar to an array of
1d-chains, each of them running along the dimer and the short bond in
the hexagonal plane.

Although such a monoclinic paramagnetic metal phase is unstable against magnetism, as we shall show in the next section, it is nevertheless of interest in view of the recent discovery of 
a high-pressure monoclinic metal phase~\cite{din14} that is actually  not dissimilar to what we just found. 
It is therefore worth investigating the mechanisms that may drive such a distortion.
\begin{figure}[ht]
\centering
(a)\includegraphics[width=0.5\columnwidth]{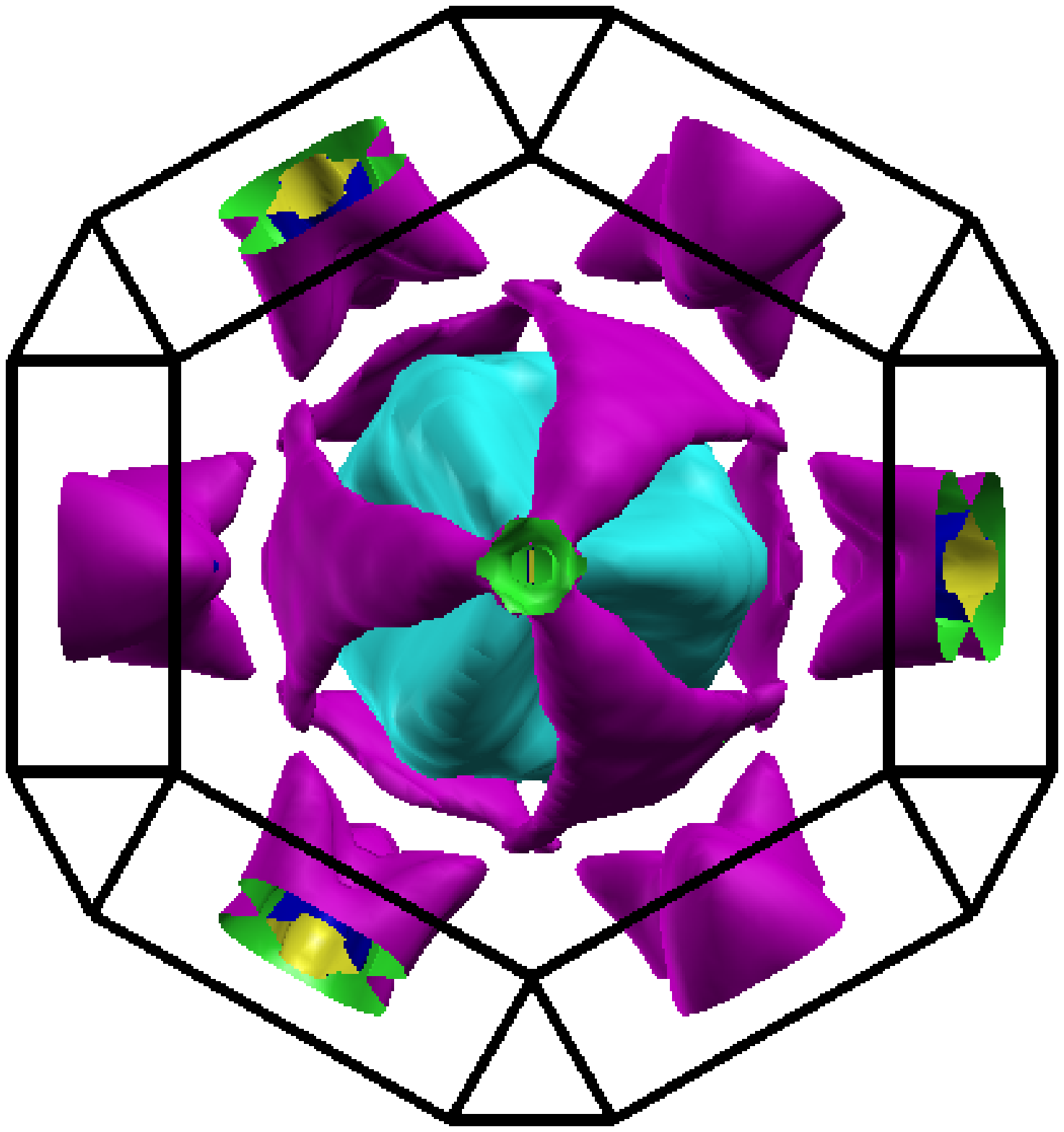}\\
(b)\includegraphics[width=0.3\columnwidth]{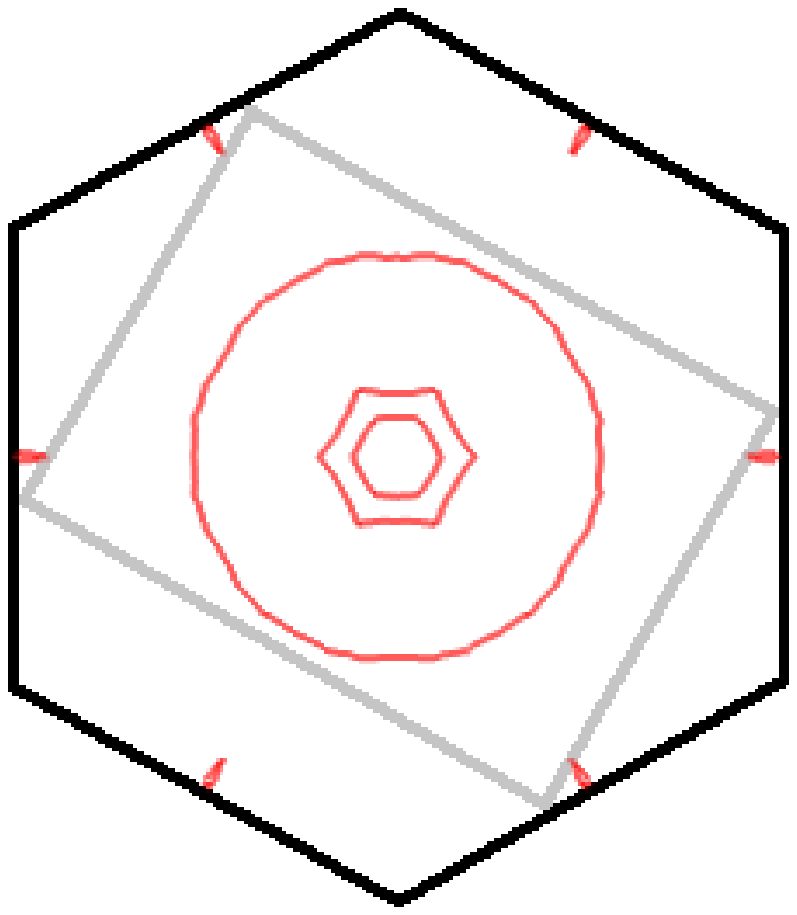}
\includegraphics[width=0.3\columnwidth]{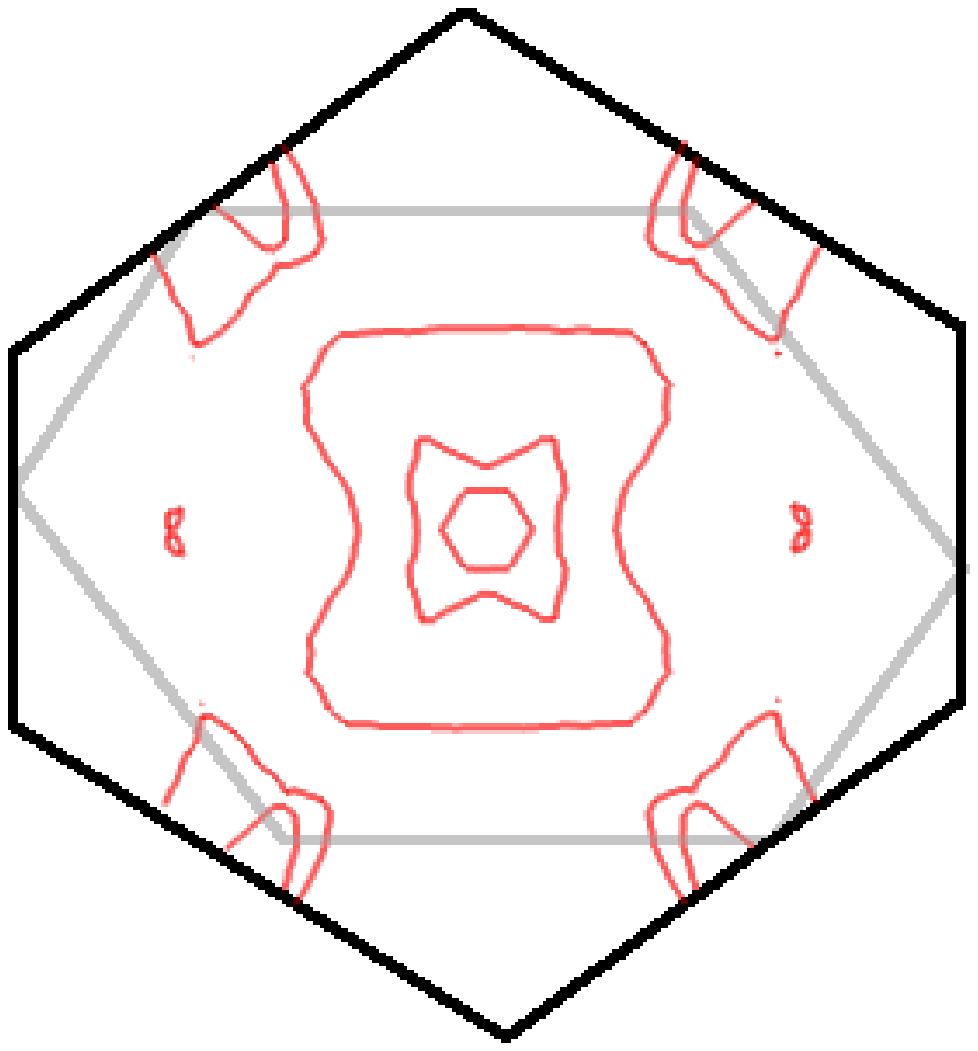}
\includegraphics[width=0.3\columnwidth]{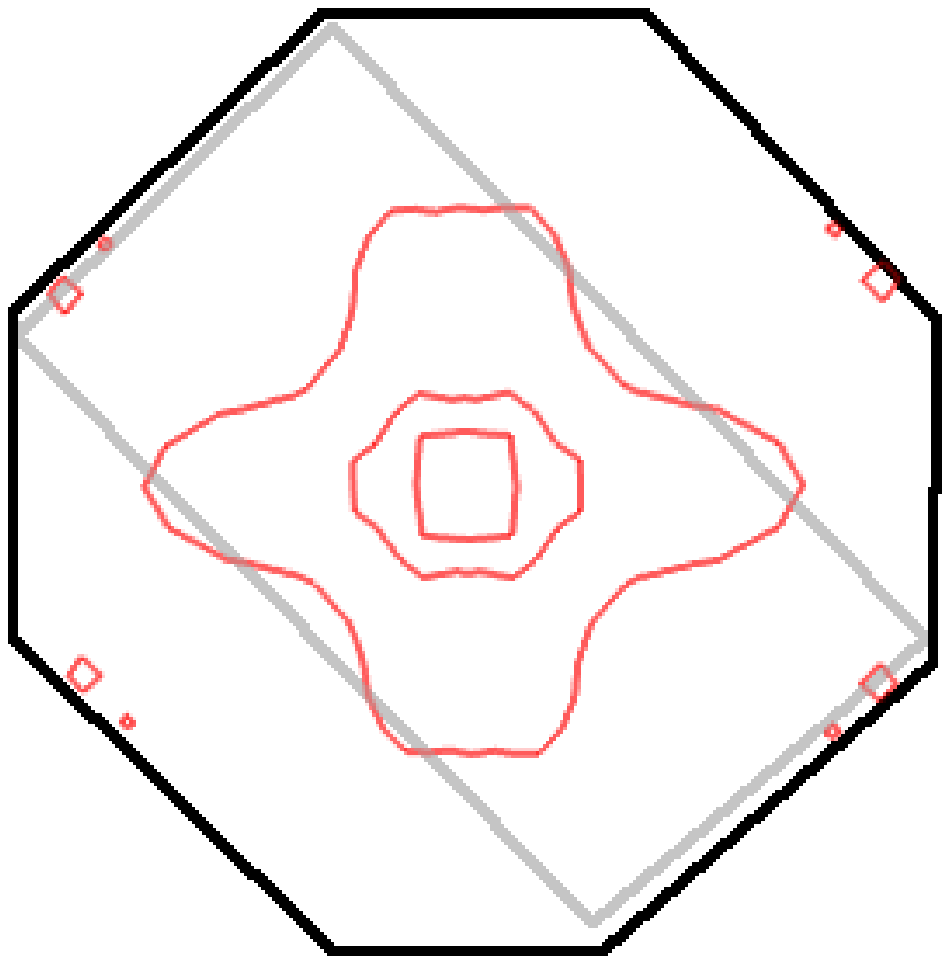}\\
\caption{\label{Fig:FermisurfaceLDAparamagnetic} (Color online) (a)
  Fermi surface of paramagnetic metallic corundum structure V$_2$O$_3$ with
  experimental atomic positions\cite{der70}, seen along the
  crystallographic $c_\text{H}$ direction, equivalently the cartesian
  $z$ axis. (b) Projections of the Fermi surface onto planes that are
  obtained by rotating around the cartesian $y$-axis of panel
  (a). While the vertical axis corresponds to the cartesian $y$
  direction, the horizontal axis is:  the cartesian $x$ direction
  (left panel); rotated about $40^\circ$ with respect to the former, so that
  it represents the direction towards a next-nearest neighbour atom in
  the adjacent hexagonal plane (middle panel); perpendicular to the
  latter (right panel). The black lines
  indicate the Brillouin zone boundary of the corundum structure, the
  grey lines its shape for the doubled monoclinic cell.}
\end{figure}

We believe that the monoclinic distortion is actually driven by a Fermi surface nesting of the corundum band structure. 
Indeed the central sheet of the whole Fermi surface, shown in
Fig.~\ref{Fig:FermisurfaceLDAparamagnetic}(a) 
and calculated with the structural data of Ref.~\onlinecite{der70} corresponding to the high-temperature paramagnetic metal, has 
nesting properties compatible with an instability towards a 
monoclinic distortion with a unit cell doubling. 
This is quite evident by looking at cut planes
as shown in
Fig.~\ref{Fig:FermisurfaceLDAparamagnetic}(b). While the
aforementioned sheet looks almost circular in the honeycomb plane of
Vanadium atoms, relatively large regions of parallel surfaces can be
identified by slightly tilting the cut-plane. 
\begin{figure}[ht]
\centering
\includegraphics[width=8.5cm]{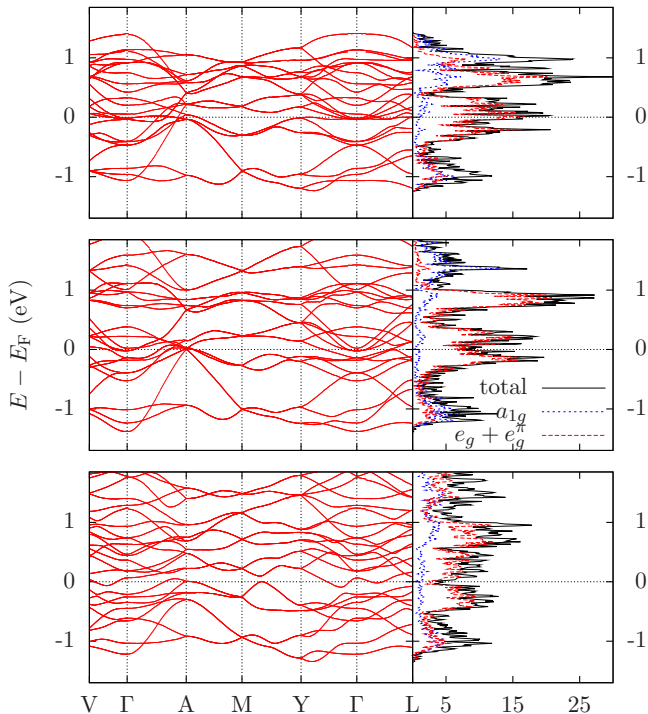}\\
\caption{\label{Fig:BDStrucPDOSLDAparamagnetic} (Color online) Band
  structure (left) and projected density of states (right) from GGA of
  paramagnetic corundum structure V$_2$O$_3$ with atomic positions determined
  experimentally for the paramagnetic metallic phase\cite{der70} (up),
  GGA-relaxed atomic positions in the corundum structure (middle) and
  GGA-relaxed atomic positions allowing for a monoclinic distortion
  (down).}
\end{figure}
In order to better uncover the driving physical mechanism, the band
structures in the enlarged 8-atom unit cell shown in
Fig.~\ref{Fig:BDStrucPDOSLDAparamagnetic} are particularly
enlightening.  The chosen path through the Brillouin zone pertaining
to the enlarged 8-atom monoclinic cell starts from the point $\mathrm
V=(0, \frac 14, -\frac 14)$ in relative reciprocal coordinates
(compatible with the magnetic order), moves further to $\Gamma$ and
along the Vanadium dimer ($\Gamma$-A direction), further goes in the
plane perpendicular thereto at the ``upper'' ($k_z = \frac 12$) edge
of the Brillouin zone (A-M), back to the original hexagonal
$a_\text{H}b_\text{H}$ plane at $k_z = 0$ (M-Y) and continues in this
plane (Y-$\Gamma$-L).

Starting from the well-known~\cite{mat94,eye05} GGA low-energy density
of states and band structure of the corundum paramagnetic metal, as
displayed in the top row of Fig.~\ref{Fig:BDStrucPDOSLDAparamagnetic},
one immediately notices a large density of states directly at the
Fermi energy, due to almost flat bands near $\Gamma$. As expected,
relaxation of the atomic positions within the corundum structure
(middle row of Fig.~\ref{Fig:BDStrucPDOSLDAparamagnetic}) partially
reduces this instability, and furthermore leads to a reduction of the
splitting between $e_g^\pi$ and high-energy $e_g$.~\cite{gri14}
However, allowing for a monoclinic distortion (bottom row of
Fig.~\ref{Fig:BDStrucPDOSLDAparamagnetic}) further stabilizes the
system by splitting the flat bands around $\Gamma$ and opening a
pseudo-gap at the Fermi energy. We mention that the undistorted
structure is not even metastable, but corresponds to a saddle point in
the total energy.


\begin{figure}[ht]
\centering
\includegraphics[width=7cm]{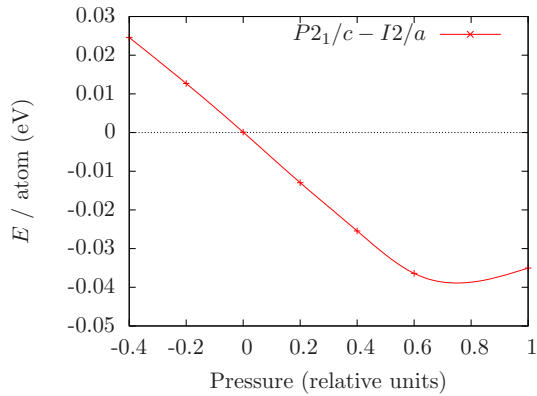}\\
\caption{\label{Fig:En_vs_Pressure} (Color online) Total energy
  difference of the $P2_1/c$ and $I2/a$ phases of monoclinic
  V$_2$O$_3$, as a function of the applied pressure. The relative
  units of the pressure axis correspond to the reported values in the
  corundum structure at ambient conditions~\cite{der70} (0.0) and the
  experimentally reported values of the high-pressure phase transition
  from $I2/a$ to $P2_1/c$~\cite{din14} (1.0).}
\end{figure}

We finally mention that the above monoclinic paramagnetic metal phase is not exactly equivalent to that observed at high-pressure in Ref.~\onlinecite{din14}, which is characterised by a further symmetry lowering from $I2/a$ to $P2_1/c$. Figure~\ref{Fig:En_vs_Pressure} shows that this symmetry reduction can also be seen in GGA upon 
simulating pressure by a decrease of the unit cell volume. 
Both phases $I2/a$ and $P2_1/c$ are minima of the total energy.  
At ambient pressure, $I2/a$ has a slightly lower energy than $P2_1/c$, but the situation is reverted already applying small pressure. Therefore the transition from $I2/a$ to $P2_1/c$ would occur according to GGA at significantly lower pressures than
reported experimentally~\cite{din14}. We believe that this is an artifact due to GGA underestimation of electronic correlations. 
Indeed, the $P2_1/c$ structure in GGA is characterised by a charge disproportionation between inequivalent Vanadium atoms, which 
is hindered by electronic correlations.

\subsection{Magnetic solutions}

\begin{table*}
  \caption{\label{Tab:GGAENERGIES} Comparison of some basic quantities, calculated for the different antiferromagnetic configurations of V$_2$O$_3$ within plain GGA. Energies are given relative to the respective ``simple'' phase.}
\begin{ruledtabular}
\begin{tabular}{lrrrr}
& ``simple'' & ``true'' & ``layered'' & para\\
\hline
experimental corundum structure\\
total GGA energy (meV/V atom) &
$0\phantom{.00}$ & -4.1 & -5.2 & 179.6\\
absolute magnetization ($\mu_{\mathrm B}$/V atom) &
1.53 & 1.57 & 1.54 & $0\phantom{.00}$\\
\hline
relaxed structures\\
total GGA energy (meV/V atom) &
$0\phantom{.00}$ & -12.2 & -7.9 & 112.1\\
absolute magnetization ($\mu_{\mathrm B}$/V atom) &
1.47 & 1.55 & 1.49 & $0\phantom{.00}$\\
\end{tabular}
\end{ruledtabular}
\end{table*}

Allowing for magnetism in the framework of spin-polarized GGA,
Tab.~\ref{Tab:GGAENERGIES} shows that the paramagnetic GGA solution
described in the previous subsection is unstable compared to all of
the magnetic solutions. Likewise, the energy gain due to its
monoclinic distortion is significantly smaller than the energy gain
due to magnetic exchange.

Comparing the different possible magnetic orderings, an important
result that we find is the stability of the ``true'' antiferromagnetic
ordering already in GGA. Note that this requires relaxation of the
atomic positions within the crystal cell. Also this relaxation of the
``true'' antiferromagnetic structure reveals a monoclinic distortion
similar to the experimentally observed one and similar (but smaller)
to the one described above for the paramagnetic regime. Note that both
the other two investigated antiferromagnetic configurations
(``layered'' and ``simple'') remain in an undistorted corundum
structure even allowing for structural relaxation. This can also be
expected by symmetry considerations of the magnetic structure,
recalling that the magnetic exchange energy gain is significantly
larger than an energy gain of the structural distortion. This
underlines the fact that the lattice distortion in antiferromagnetic
V$_2$O$_3$ directly stabilizes the antiferromagnetic order, and is thus
closely related to it.

\begin{figure}[th]
\centering
\includegraphics[width=8.5cm]{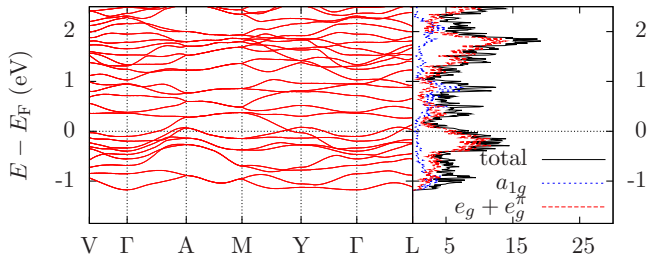}\\
\caption{\label{Fig:BDStrucPDOSLDA} (Color online) Band structure
  (left) and projected density of states (right) of true
  antiferromagnetic V$_2$O$_3$ from GGA.}
\end{figure}
As can be seen from the low-energy density of states and band
structure shown in Fig.~\ref{Fig:BDStrucPDOSLDA}, the main drawback is
that pure GGA cannot describe the insulating behavior of the magnetic
structure. This is the expected result in view of the importance of
strong electronic correlations in V$_2$O$_3$. Furthermore, one notices
that the splitting between $a_{1g}$ and $e_g^\pi$ turns out to be
small enough that all orbitals from the $t_{2g}$ block have a similar
filling (with the $a_{1g}$ occupation slightly smaller than each
$e_g^\pi$). This gives rise to a spin magnetic moment slightly smaller
than the observed experimental value, which we recall is $\sim 1.7
\mu_B$ once corrected by the orbital contribution.

\section{DFT+U studies\label{Sec:DFT+U}}

Since the antiferromagnetic insulating state is the ground state of
V$_2$O$_3$, one can expect to gain insight into the effects of strong
electronic correlations already from a simple method like
GGA+U. Here, the simplified version of Cococcioni and
de~Gironcoli~\cite{coc05} in the {\sc Quantum ESPRESSO} package is put
into practice, which implies the use of only one effective parameter
$U-J$ and a fully-localized limit (FLL) double-counting correction.

\begin{figure}[ht]
\centering
\includegraphics[width=7cm]{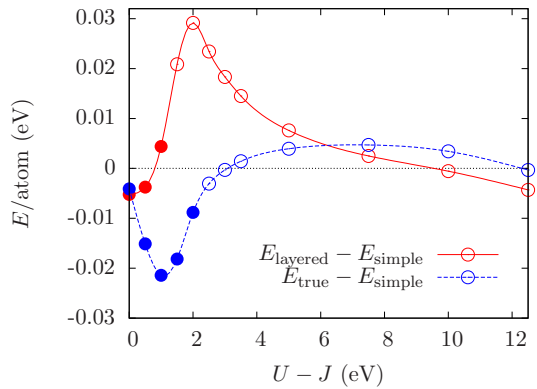}\\
\caption{\label{Fig:EofULDAUortho} (Color online) Phase stability
  comparison in the experimentally determined corundum structure of
  V$_2$O$_3$ within GGA+U. Closed circles indicate metallic, open
  circles indicate insulating solutions.}
\end{figure}

Since the actual value of the parameter $U-J$ is a priori unknown,
Fig.~\ref{Fig:EofULDAUortho} compares the stability of each of the
possible magnetic structures in terms of their total energy for a
wide parameter range. This calculation is performed in the experimental
atomic positions in the unrelaxed corundum structure,\cite{der70} nevertheless already shows that a small
$U-J$ is able to stabilize the ``true'' antiferromagnetic
phase, whereas for large interactions the ``simple'' and ``layered'' structures 
become more stable. 


\begin{figure}[ht]
\centering
\includegraphics[width=7cm]{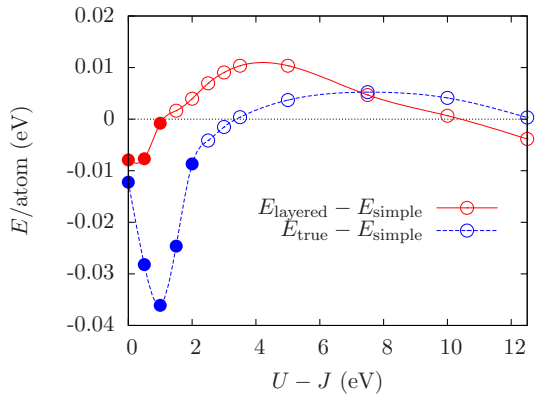}\\
\caption{\label{Fig:EofULDAUrelaxed} (Color online) Phase stability
  comparison in the relaxed structure of V$_2$O$_3$ within
  GGA+U. Filled circles indicate metallic solutions of the
  layered/true phase, open circles indicate insulating solutions
  thereof.}
\end{figure}

\begin{figure}[ht]
\centering
\includegraphics[width=8.5cm]{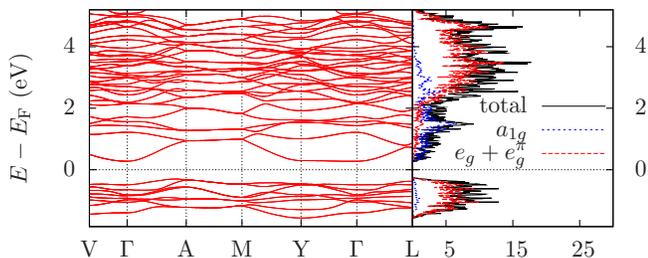}\\
\caption{\label{Fig:BDStrucPDOSLDAU3} (Color online) Band structure
  (left) and projected density of states (right) of true
  antiferromagnetic V$_2$O$_3$ from GGA+U at $U-J=3.0$~eV.}
\end{figure}
The relaxation of the structural parameters within GGA+U further stabilizes the ``true''
antiferromagnetically ordered phase and extends its stability region, as  shown
in Fig.~\ref{Fig:EofULDAUrelaxed}. As anticipated, above $U-J\simeq 2.0$~eV, the "true" antiferromagnetic state turns from metallic into insulating, which 
sets between 2.0 and 4.0~eV the range of $U-J$ values that reproduce within GGA+U the 
magnetic and conducting properties of the actual material. 
Fig.~\ref{Fig:BDStrucPDOSLDAU3} shows the
GGA+U Kohn-Sham band structure and density of states projected onto
Vanadium $d$-orbitals for $U-J = 3.0$~eV, which we assume to be a realistic estimate. 

It turns out that the GGA+U realization of the insulating phase corresponds to the straightforward
solution of occupied $e^\pi_g$ orbitals and empty $a_{1g}$ one, 
thus implying an $S=1$ spin configuration. As
mentioned, this is in line with e.~g. the $U$-induced paramagnetic
Mott metal-insulator transition in DMFT\cite{pot07}, though still a matter of
debate.~\cite{mat05,mat09}



\begin{figure}[ht]
\centering
\includegraphics[width=7cm]{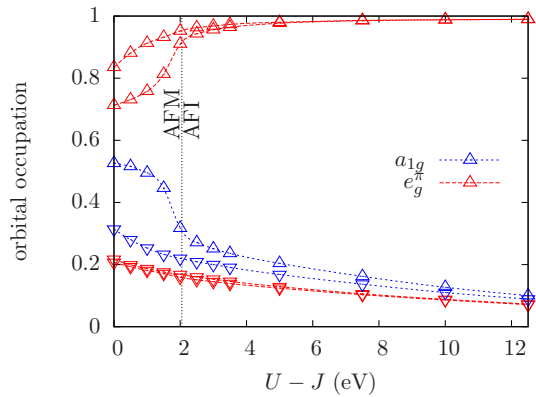}\\
\caption{\label{Fig:EVofULDAU} (Color online) Crystal field basis
  occupation numbers (eigenvalues of the GGA+U local density matrix)
  per Vanadium d-orbital as a function of $U-J$ in the relaxed true
  antiferromagnetic phase of V$_2$O$_3$. Triangles pointing up
  indicate the majority spin channel, triangles pointing down the
  minority spin channel. As before, AFM and AFI stand for
  antiferromagnetic metal and insulator, respectively.}
\end{figure}
%
This scenario is further confirmed in Fig.~\ref{Fig:EVofULDAU}, where we plot the occupation numbers in the crystal field basis that diagonalizes the GGA+U occupation
matrix. We mention that, since these orbitals are not strongly localized
on a single Vanadium atom, the total occupation is not precisely 2.
The depletion of the $a_{1g}$-like crystal-field basis orbital with
increasing values of $U-J$ in favor of $e^\pi_g$-like orbitals is evident. 
A noteworthy jump occurs at the metal-insulator
transition, with the above-mentioned scenario of a majority-spin
$a_{1g}$ orbital that is almost half-filled below and almost empty
above. 


One remarkable effect of the monoclinic distortion and the magnetic
ordering is the lifting of the degeneracy bewteen the two $e^\pi_g$
orbitals, which is particularly strong in the (unphysical) magnetic
metallic regime, but still present in the magnetic insulating
one. It is a direct consequence of the breaking of the three-fold
rotational symmetry of the Vanadium planes with magnetic order or
structural distortion, and therefore does not occur for the other
investigated magnetic structures that do not break such symmetry. 

%
\begin{figure}[ht]
\centering
\includegraphics[width=7cm]{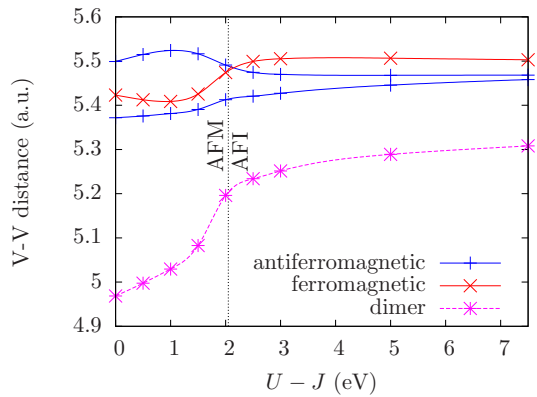}\\
\caption{\label{Fig:AtomicDistancesLDAUrelaxed} (Color online) Nearest-neighbour Vanadium distances from relaxation in the ``true'' antiferromagnetic structure, indicating the monoclinic distortion.
AFM and AFI stand for antiferromagnetic metal and insulator, respectively.}
\end{figure}
The inequality between the  two $e^\pi_g$ orbitals, more accentuated the smaller $U-J$, 
also shows up into different lengths of the two antiferromagnetic in-plane nearest-neighbor bonds, 
as shown in Fig.~\ref{Fig:AtomicDistancesLDAUrelaxed}, to such an extent that, in the metal phase 
$U-J < 2$~eV, one of the antiferromagnetic bonds is the longest. On the contrary, 
above $U-J = 2$~eV, i.e. in the realistic insulating phase,  all nearest-neighbor Vanadium distances 
shown in the same Fig.~\ref{Fig:AtomicDistancesLDAUrelaxed} 
are compatible with the observed monoclinic distortion and 
with the intuitive expectation of elongated ferromagnetic bonds and shortened 
antiferromagnetic ones.  However a small difference between the 
lengths of the two in-plane antiferromagnetic bonds persists also in the insulating side. 


\section{Magnetic anisotropy\label{Sec:Orient}}

Already the original 1970's work by Moon~\cite{moo70} pointed out that the
magnetic moments are oriented with a certain angle towards the
crystallographic $c$-direction. Such a
magnetic anisotropy is, at first glance, not expected for a light
element like Vanadium. However, for similar compounds like Vanadium
spinels, values of spin-orbit coupling in the range of 13-20~meV have
been reported\cite{abr70,col70,tch04,per07}, which makes a DFT
calculation including relativistic effects (spin-orbit coupling) and
non-collinear magnetism worth trying. To this end, the implementation
thereof\cite{cor05} in {\sc Quantum ESPRESSO} has been used with a
fully relativistic pseudopotential
(V.rel-pbe-spnl-rrkjus\_psl.1.0.0.UPF from the PSLibrary of
http://www.quantum-espresso.org) and the rotationally invariant GGA+U
formulation of Liechtenstein et~al.\cite{lie95}. Due to the high
computational demands of such a calculation, no further relaxation of
the structural parameters has been done, but values of the
monoclinically distorted collinear-antiferromagnetic structure relaxed
without $U$ have been used.

\begin{figure}[ht]
\centering
\includegraphics[width=7cm]{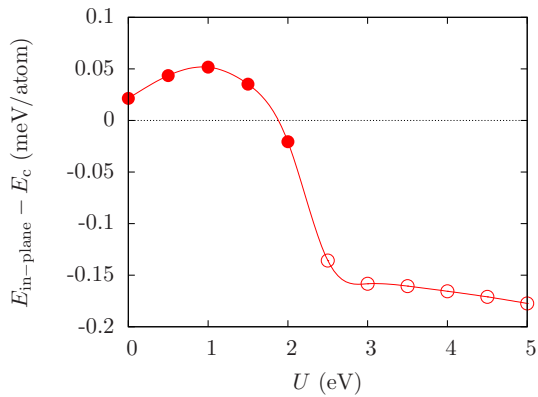}\\
\caption{\label{Fig:EnoncollinofULDAU} (Color online) LDA+U energy
  difference of ``true'' antiferromagnetic solutions with spin
  orientation in the honeycomb plane and in $c$-direction
  (perpendicular thereto), as a function of the Hubbard parameter
  $U$. Closed circles indicate metallic, open circles indicate
  insulating solutions.}
\end{figure}

Fig.~\ref{Fig:EnoncollinofULDAU} shows the energy difference between
GGA+U solutions with magnetic moments oriented in the honeycomb plane
($E_{\mathrm{in-plane}}$) and oriented perpendicular
thereto ($E_{\mathrm c}$), i.~e. in the crystallographic
$c$-direction. A first result is that, without electronic
correlations, i.~e. at the GGA level, the $c$-axis alignment  is
favored, although with a tiny energy difference. However, upon increasing $U$, 
and specifically above the metal-insulator transition, 
the in-plane orientation becomes more stable. 

Indeed there is a close connection between the change from easy-axis to easy-plane and 
the $a_{1g}$ depletion that occurs at the metal-insulator 
transition, which can be illustrated by a simple model calculation. 

Let us assume an isolated Vanadium in the trigonal crystal field. 
A suitable basis set for the $t_{2g}$ manifold (with corundum
symmetry) can be written (the sign is related to the multiple Vanadium atoms per unit cell):
\begin{eqnarray}
\vert a_{1g} \rangle &=& \vert d_{3z^2-r^2} \rangle, \nonumber\\
\label{Eqn:SINGLEPARTICLEBASIS}
\vert e_{g1} \rangle &=& \sqrt \frac 23 \, \vert d_{xy} \rangle \pm \sqrt \frac 13 \, \vert d_{xz} \rangle,\\
\vert e_{g2} \rangle &=& - \sqrt \frac 23 \, \vert d_{x^2-y^2} \rangle \mp \sqrt \frac 13  \,\vert d_{yz} \rangle. \nonumber
\end{eqnarray}
Using the representation $\vert n_{a_{1g}}, n_{e_{g1}}, n_{e_{g2}}; S_z
\rangle$ to denote two electrons coupled into a spin triplet with $z$-component $S^z=-1,0,1$ 
and  occupying the 
single-particle states Eq.~\eqn{Eqn:SINGLEPARTICLEBASIS} with occupation $n_i=0,1$ such that  
$\sum n_i = 2$, we define new states:
\begin{eqnarray}
\vert 1, S_z \rangle &=& \mp \sqrt{ \frac{1}{2}}\; \Big(\vert 1,0,1;S_z \rangle - i \vert 1,1,0 ;S_z\rangle\Big), \nonumber\\
\vert 0, S_z \rangle &=& \vert 0,1,1;S_z \rangle,\label{L-basis}\\
\vert -1, S_z \rangle &=& \mp \sqrt{ \frac{1}{2}}\; \Big( - \vert 1,0,1;S_z \rangle - i \vert 1,1,0 ;S_z\rangle
\Big), \nonumber
\end{eqnarray}
which are actually eigenstates of the $z$-component of the angular momentum operator $L_z$ 
projected onto the $t_{2g}$ manifold that effectively realizes an $l=1$ representation, 
i.e. $L_z\,\vert M,S_z\rangle  = M\,\vert M,S_z\rangle$, with $M=-1,0,1$. 
The spin-orbit coupling projected onto the basis Eq.~\eqn{L-basis} reads  
\be
H_\text{SOC} = \fract{\lambda_\text{SOC}}{2}\,\Big(2\,S_z\,L_z 
+ S^+\,L^+ + S^-\,L^-\Big),\label{SOC}
\ee
where $L^\pm$ have the same expression as for $l=1$ angular momentum operators, while 
the trigonal crystal field can be written as 
\be
H_\text{tr} = 3 V_\text{tr}\,\bigg(L_z^2-\frac{2}{3}\bigg).\label{CF}
\ee
One can easily realize that, for $V_\text{tr}=0$, the lowest energy state at $\lambda_\text{SOC}\not = 0$ 
is five-fold degenerate, corresponding to two $d$-electrons coupled according to the Hund's rules 
to $S=1$, $L=3$ and $J=2$. Conversely, for $\lambda_\text{SOC} = 0$ but $V_\text{tr}\not = 0$, 
the lowest energy state $\vert 0,S_z\rangle$ is an orbitally non-degenerate spin-triplet. 

However, when both parameters are finite with $\lambda_\text{SOC}\ll V_\text{tr}$, 
the lowest energy state is 
\be
\vert 0 \rangle \equiv \cos\theta\,\vert0,0\rangle - \fract{\sin\theta}{\sqrt{2}}\;\Big(
\vert 1,1 \rangle+\vert -1,-1 \rangle\Big),
\ee
with 
\[
\tan 2\theta = \fract{2\sqrt{2}\;\lambda_\text{SOC}}{3V_\text{tr} + \lambda_\text{SOC}},
\]
followed by the doublet
\be
\vert \pm 1 \rangle \equiv \cos\phi\,\vert 0,\pm 1\rangle - \sin\phi\,\vert \mp 1,0\rangle,
\ee
with $\tan 2\phi = 2\lambda_\text{SOC}/3V_\text{tr}$. 

If we regard the three states 
$\vert 0\rangle$ and $\vert \pm 1\rangle$ as the effective $S=1$ states of each isolated Vanadium, 
the above results show that the spin-orbit coupling generates an easy-plane anisotropy
\be
H_* = \Gamma_\text{tr}\, S_z^2 \,\simeq \, \fract{\lambda^2_\text{SOC}}{3V_\text{tr}}\, S_z^2.
\label{H*-1}
\ee

We could proceed and consider the effects of a monoclinic distortion $V_m\ll V_\text{tr}$ that makes the $e^\pi_g$ orbitals inequivalent.\cite{tan02} Looking at the basis set Eq.~\eqn{L-basis}, one realizes that the leading effect of this distortion corresponds to the additional operator 
\be
H_m = V_m\,\Big(L^z L^y + L^y L^z\Big),
\ee
whose action on the $S=1$ basis $\vert 0\rangle$ and $\vert \pm 1\rangle$ above is equivalent to an 
additional anisotropy term
\be
\delta H_* = \Gamma_m\, \Big(S^z S^y + S^y S^z\Big),\label{H*-2}
\ee
which, together with Eq.~\eqn{H*-1}, might indeed justify the magnetic moment lying in the monoclinic $a$-$c$ plane, assumed here to be the $y$-$z$ plane, $29^\circ$ above the hexagonal basal plane. 

Evidently, the above calculation is a very rough one. However it shows that one can rationalize the observed magnetic anisotropy already at the level of a  single Vanadium, without  invoking the "dimer" as a building block.\cite{tan02}

\section{Model studies and Discussion\label{Sec:Models}}

In the previous sections we have shown that the GGA+U accurately accounts for  
the magnetic, structural and conducting properties of V$_2$O$_3$ in its low temperature phase. 
This suggests the possibility of recovering similar results by model calculations within an independent-particle approximation. 

An obvious starting point is a three-orbital
tight-binding  model for the corundum phase supplied by a local Hubbard $U$ and a Coulomb exchange $J$. 
\begin{table}
  \caption{\label{Tab:TBParameters} Tight-binding parameters (in eV) obtained for corundum phase V$_2$O$_3$ by Saha-Dasgupta~{\it et~al.}~\cite{sah09} Nomenclature according to Ref.~\onlinecite{cas78_2}.}
\begin{ruledtabular}
\begin{tabular}{cccccc}
$\mu$     & 
$\rho$    & 
$-\alpha$ & 
$\beta$   & 
$\sigma$  & 
$-\tau$   \\
\hline
$ 0.06$ &
$-0.51$ &
$ 0.08$ &
$-0.21$ &
$-0.03$ &
$-0.26$ \\
\end{tabular}
\end{ruledtabular}
\end{table}
To allow comparison with previous works, tight-binding parameters
calculated by Saha-Dasgupta~{\it et~al.}~\cite{sah09} in an \textit
NMTO basis set are used, which also reproduce well our GGA band structure. They are summarized in
Tab.~\ref{Tab:TBParameters}, following the nomenclature of
Ref.~\onlinecite{cas78_2}. In comparison with the parameters obtained by
Refs.~\onlinecite{cas78_2} and \onlinecite{mat94}, they are characterized by
relatively small out-of-plane hopping amplitudes, $\mu$ and $\rho$, and by slightly different 
in-plane values, $\alpha$ and $\beta$. According to
Ref.~\onlinecite{sah09}, the monoclinic distortion would lead to a
change up to 4~\% of the hopping amplitudes, with the possible 
exception of the out-of-plane ones. We shall therefore not account for those changes and keep using the corundum structure parameters. The two-particle interaction, which  
includes a Hubbard $U$ and a Coulomb exchange that we assume equal to $J=U/4$, 
is treated within the Hartree-Fock approximation. This amounts to consider a trial wavefunction 
ground state of a non-interacting Hamiltonian with the same hopping amplitudes and, in addition, 
with spin and orbital dependent on-site energies. We shall use an 8-site supercell, which allows to 
describes the "true" antiferromagnetic state, so that there are in principle 8 sets of Hartree-Fock on-site energies to be determined by minimizing the average total energy. 


\begin{figure}[ht]
\centering
\includegraphics[width=7cm]{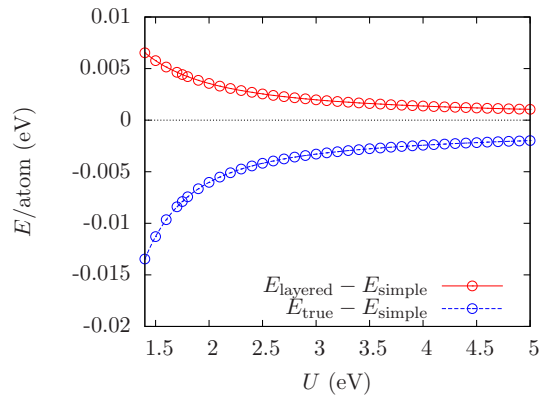}
\caption{\label{Fig:EofUTB} (Color online) Phase stability of the
  insulating phases from a Hartree-Fock calculation using tight
  binding parameters obtained by Saha-Dasgupta~{\it
    et~al.}~\cite{sah09} for corundum V$_2$O$_3$.}
\end{figure}

Fig.~\ref{Fig:EofUTB} shows the Hartree-Fock total energies 
for each of the magnetic configurations 
that are obtained by a suitable choice of the initial
configuration for values of $U$ above which all solutions are insulating. 
The comparison with GGA+U in the corundum phase is in fact not bad, although we cannot 
compare directly the value of $U$ used here with that in GGA+U, which already at $U=0$ includes 
correlation effects. Therefore it is not surprising that the "true" structure remains the lowest energy one 
at least up to $U=5$. 

\begin{figure}[ht]
\centering
\includegraphics[width=7cm]{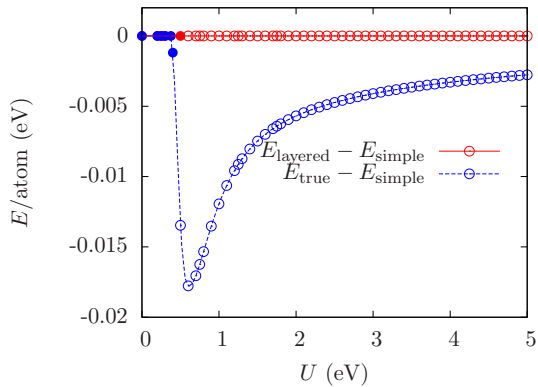}
\caption{\label{Fig:EofUTBdelta40} (Color online) Phase stability from
  a Hartree-Fock calculation using tight binding parameters obtained
  by Saha-Dasgupta~{\it et~al.}~\cite{sah09} for corundum V$_2$O$_3$,
  but fixing all hopping perpendicular to the Vanadium planes to
  0. Closed circles indicate metallic, open circles indicate
  insulating solutions.}
\end{figure}

Bearing in mind the comparatively large differences of the reported
values of the out-of-plane hopping, a first check of the model accuracy is to estimate
the influence thereof onto the overall phase stability,
which is highlighted in Fig.~\ref{Fig:EofUTBdelta40} in the extreme limit 
of vanishing out-of-plane hopping parameters, $\mu=\rho=0$. 
This check is also related to one of the main effects of the
monoclinic distortion, which tilts and stretches the out-of-plane
Vanadium dimer bonds, see 
Fig.~\ref{Fig:AtomicDistancesLDAUrelaxed}. When $\mu=\rho=0$, the
``layered'' and ``simple'' orderings are obviously degenerate, but the ``true''
one is still stable up to large values of $U$, although with a smaller energy difference. This suggests that
the reason for the stability of the ``true'' structure is 
primarily in the in-plane physics. The out-of-plane hopping
further stabilizes this phase, at the same time destabilizing
the ``layered'' structure compared to the "simple" one.

\begin{figure}[ht]
\centering
\includegraphics[width=7cm]{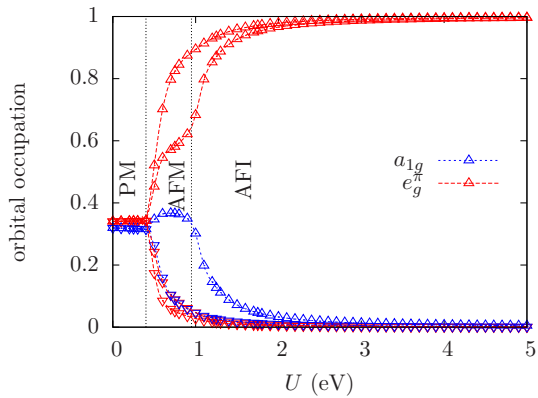}
\caption{\label{Fig:DenmatLocHF} (Color online) Orbital occupations in
  the ``true'' antiferromagnetic phase from Hartree-Fock as a function
  of $U$. Triangles pointing up indicate the majority spin channel,
  triangles pointing down the minority spin channel. PM, AFM and AFI
  stand for paramagnetic metal, antiferromagnetic metal and
  antiferromagnetic insulator, respectively.}
\end{figure}

The occupation numbers of the three orbitals on each Vanadium site,
displayed in Fig.~\ref{Fig:DenmatLocHF}, show a very similar
behavior to the GGA+U results of Fig.~\ref{Fig:EVofULDAU}, with the
exception of the small-$U$ paramagnetic metal phase. In the intermediate
(unphysical) antiferromagnetic metallic regime, one can even see a
slight increase of the majority-spin $a_{1g}$ occupations, but a sharp
decrease at the metal-insulator transition reveals again the scenario in which 
the two electrons per site occupy the $e_g^\pi$ orbitals, making the 
$a_{1g}$ orbitals practically empty. Note that this orbital
polarization turns out to be stronger than in GGA+U, which can be
attributed to the fact that the total occupation of the Hartree-Fock
model is kept fixed at 2, so that no contributions of
e.~g. neighboring oxygen atoms can occur. How to adapt the occupation
numbers to compare with experimental findings is shortly discussed
in Ref.~\onlinecite{pot07}. 

Furthermore, the aforementioned splitting of the
$e_g^\pi$ orbitals shows up also in the ``true'' antiferromagnetically
ordered phase with hopping parameters of the corundum structure, as
opposed to the ``layered'' and ``simple'' ordering. This is a further
evidence that the structural distortion is intimately tight to the magnetic ordering. We highlight that such a splitting is uniform within all the eight sites of the supercell, as we also found by GGA+U. In other words, both Hartree-Fock approximation and GGA+U predict a ferro-orbital ordering within the $e^\pi_g$ manifold, unlike the antiferro-orbital one found in Ref.~\onlinecite{cas78_2}.  

\begin{figure}[ht]
\centering
\includegraphics[width=7cm]{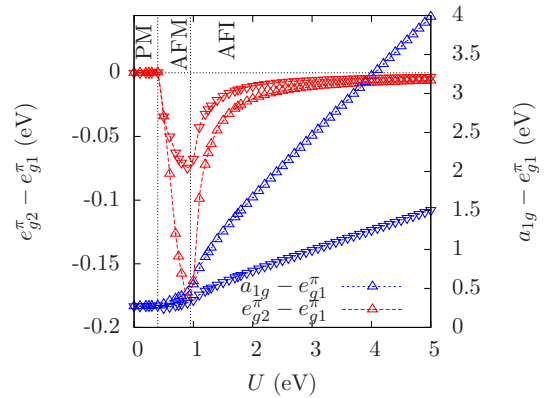}
\caption{\label{Fig:OnsiteHF} (Color online) Differences of the
  on-site energy terms in the ``true'' antiferromagnetic phase from
  Hartree-Fock as a function of $U$. Triangles pointing up indicate
  the majority spin channel, triangles pointing down the minority spin
  channel.}
\end{figure}

The Hartree-Fock on-site energy terms, shown in
Fig.~\ref{Fig:OnsiteHF}, display the well-known large increase of
the $e_g$-$a_{1g}$ crystal field splitting with increasing value of
$U$, especially for the majority spin, whereas all three orbitals are
basically unoccupied for the minority spin. Also the splitting of the
$e_g$ orbitals is visible, which amounts to the relatively large
value of 30~meV for the majority spin in the region of realistic parameter values. We note that the lowering of $e^\pi_{g1}$ with respect to $e^\pi_{g2}$ follows from our choice of a specific "true" magnetic order among the three equivalent ones allowed by the original $C_3$ symmetry.  

\begin{figure}[ht]
\centering
\includegraphics[width=6.5cm]{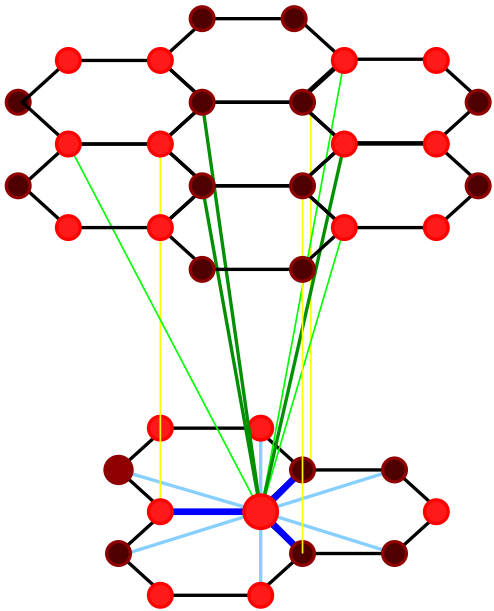}
\caption{\label{Fig:AFIbondSketch} (Color online) Illustration of the
  magnetic ordering (different spin directions as light red 
  and dark red) and of the different relevant hopping
  processes from the central atom shown as a large circle. Besides the nearest neighbour $e^\pi_g-e^\pi_g$ direct hopping, dark blue, 
  we draw the $a_{1g}$-mediated next nearest neighbour ones: light blue in the hexagonal plane, and green between planes. Bold green lines denote that two independent paths produce the same hopping process.}
\end{figure}

In conclusion, even the Hartree-Fock approximation to a three-band Hubbard model with realistic hopping parameters reproduces the correct magnetic structure and indicates the tendency towards a spontaneous monoclinic distortion. 

\subsection{Role of the $a_{1g}$ orbital}

Within both GGA+U and Hartee-Fock the effect of $U$ is to increase repulsion between occupied and unoccupied states, hence between 
majority and minority spins and between $e^\pi_g$ and $a_{1g}$ orbitals. Therefore, at large enough $U$, the $a_{1g}$ orbital can be discarded and one expects a two-sublattice antiferromagnetic order to prevail. This is indeed the case, see 
Fig.~\ref{Fig:EofULDAUrelaxed}. It thus follows that the $a_{1g}$ orbital must play a role to stabilise the "true" magnetic ordering for intermediate values of $U$. We argue that such a role is to provide magnetic frustration. 

Let us imagine that the effective crystal field splitting, enhanced by $U$, is large enough that we can treat the hopping $\tau$, see Table~\ref{Tab:TBParameters}, between 
$e^\pi_g$ and $a_{1g}$ in perturbation theory. Let us focus on the Vanadium atom drawn as a large circle in Fig.~\ref{Fig:AFIbondSketch}. At second order, $\tau$ induces next nearest neighbour $e^\pi_g-e^\pi_g$ hopping terms, shown as light blue lines in Fig.~\ref{Fig:AFIbondSketch}, that compete against the nearest neighbour ones, drawn in dark blue. If we also take into account the large direct $a_{1g}-a_{1g}$ hopping along the $c$-axis, $\rho$ in 
Table~\ref{Tab:TBParameters}, next nearest neighbour 
$e^\pi_g-e^\pi_g$ hopping processes between adjacent planes are   
generated, see green lines in Fig.~\ref{Fig:AFIbondSketch}, where the bold ones indicate that there are two different paths contributing to  that process.
 
If we could use these hopping elements to derive an effective $S=1$ Heisenberg model, we would find on each plane both nearest, $J_1$, and next nearest, $J_2$, neighbour exchange constants. 
The phase diagram of the $S=1/2$ $J_1$-$J_2$ Heisenberg model
on the honeycomb lattice is relatively well known.~\cite{fou01,alb11,mez12} The two-sublattice antiferromagnet is stable 
for $J_2\lesssim 0.2 J_1$. For $0.2\lesssim J_2/J_1 \lesssim 0.4$, 
there seems to be no magnetic order. Finally, for $J_2\gtrsim 0.4 J_1$ the magnetic order is exactly the "true" antiferromagnet, 
shown in Fig.~\ref{Fig:AFIbondSketch}. We cannot exclude that the larger spin $S=1$ and the coupling between the planes could stabilise in our case the "true" antiferromagnet even in the formerly disordered region $0.2\lesssim J_2/J_1 \lesssim 0.4$. 

\begin{figure}[th]
\centering
\includegraphics[width=7cm]{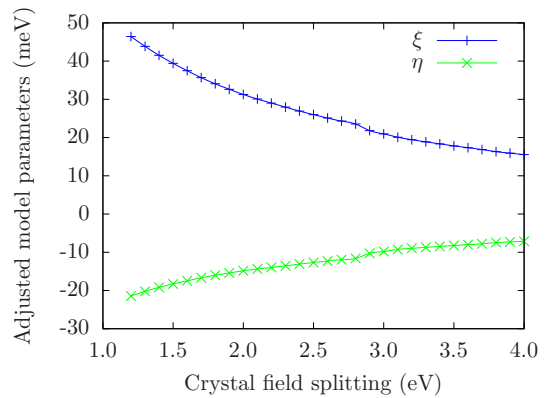}
\caption{\label{Fig:XIofA1g} (Color online) Fitted next-nearest
  neighbour in-plane hopping parameters $\xi$ and $\eta$. See text for details.}
\end{figure}
In order to estimate the amount of frustration brought by the $a_{1g}$-mediated next-nearest neighbour hopping, we have calculated the  band structure in the corundum phase as function of an artificial crystal field spitting. 
When the crystal field is large enough, $\gtrsim 1.2$~eV, the upper $a_{1g}$-derived bands are well separated from the lower $e^\pi_{g}$-derived ones, hence we can fit the latter by a simple tight-binding model with only 
$e^\pi_{g}-e^\pi_{g}$ nearest and next-nearest neighbour hopping parameters. The nearest neighbour ones 
are assumed to be those in Table~\ref{Tab:TBParameters}. 
By the three-fold rotational symmetry, we just need two fitting next-nearest neighbour hopping parameters within the hexagonal planes, which we call $\xi$ and $\eta$ and plot in Figure~\ref{Fig:XIofA1g}. 
As expected, $\xi$ and $\eta$ decrease in absolute value as 
the crystal field increases. However, for not too large crystal field splitting, $\xi$ and $\eta$ have the same order of magnitude as 
the nearest neighbour hopping parameters $\alpha$ and $\beta$, see 
Table~\ref{Tab:TBParameters}. 

The above calculation is very rough; the effective crystal field is not so large that $e^\pi_g$ and $a_{1g}$ bands are separated, and $U$ not big enough to justify a mapping onto an Heisenberg model. 
However we believe that the overall scenario is correct: the 
$a_{1g}$ orbital, although pushed by relatively strong correlations above Fermi in the insulating phase, as first noted by DMFT in Ref.~\onlinecite{pot07}, still heavily contributes to stabilise the "true" magnetic structure. All these features, including the monoclinic distortion and the magnetic anisotropy, are well captured by an independent particle scheme as GGA+U.

\begin{acknowledgments}
  We wish to thank S. Shahab Naghavi, Ryan Tyler Requist, Matteo
  Sandri, Andrea dal Corso and Erio Tosatti for very helpful
  discussions. This work has been supported by the European Union,
  Seventh Framework Programme, under the project GO~FAST, grant
  agreement no.~280555.
\end{acknowledgments}

\bibliography{bibd}

\begin{thebibliography}{42}
\expandafter\ifx\csname natexlab\endcsname\relax\def\natexlab#1{#1}\fi
\expandafter\ifx\csname bibnamefont\endcsname\relax
  \def\bibnamefont#1{#1}\fi
\expandafter\ifx\csname bibfnamefont\endcsname\relax
  \def\bibfnamefont#1{#1}\fi
\expandafter\ifx\csname citenamefont\endcsname\relax
  \def\citenamefont#1{#1}\fi
\expandafter\ifx\csname url\endcsname\relax
  \def\url#1{\texttt{#1}}\fi
\expandafter\ifx\csname urlprefix\endcsname\relax\def\urlprefix{URL }\fi
\providecommand{\bibinfo}[2]{#2}
\providecommand{\eprint}[2][]{\url{#2}}

\bibitem[{\citenamefont{Poteryaev et~al.}(2007)\citenamefont{Poteryaev,
  Tomczak, Biermann, Georges, Lichtenstein, Rubtsov, Saha-Dasgupta, and
  Andersen}}]{pot07}
\bibinfo{author}{\bibfnamefont{A.~I.} \bibnamefont{Poteryaev}},
  \bibinfo{author}{\bibfnamefont{J.~M.} \bibnamefont{Tomczak}},
  \bibinfo{author}{\bibfnamefont{S.}~\bibnamefont{Biermann}},
  \bibinfo{author}{\bibfnamefont{A.}~\bibnamefont{Georges}},
  \bibinfo{author}{\bibfnamefont{A.~I.} \bibnamefont{Lichtenstein}},
  \bibinfo{author}{\bibfnamefont{A.~N.} \bibnamefont{Rubtsov}},
  \bibinfo{author}{\bibfnamefont{T.}~\bibnamefont{Saha-Dasgupta}},
  \bibnamefont{and} \bibinfo{author}{\bibfnamefont{O.~K.}
  \bibnamefont{Andersen}}, \bibinfo{journal}{Phys. Rev. B}
  \textbf{\bibinfo{volume}{76}}, \bibinfo{pages}{085127}
  (\bibinfo{year}{2007}).

\bibitem[{\citenamefont{McWhan et~al.}(1971)\citenamefont{McWhan, Remeika,
  Rice, Brinkman, Maita, and Menth}}]{mcw71}
\bibinfo{author}{\bibfnamefont{D.~B.} \bibnamefont{McWhan}},
  \bibinfo{author}{\bibfnamefont{J.~P.} \bibnamefont{Remeika}},
  \bibinfo{author}{\bibfnamefont{T.~M.} \bibnamefont{Rice}},
  \bibinfo{author}{\bibfnamefont{W.~F.} \bibnamefont{Brinkman}},
  \bibinfo{author}{\bibfnamefont{J.~P.} \bibnamefont{Maita}}, \bibnamefont{and}
  \bibinfo{author}{\bibfnamefont{A.}~\bibnamefont{Menth}},
  \bibinfo{journal}{Phys. Rev. Lett.} \textbf{\bibinfo{volume}{27}},
  \bibinfo{pages}{941} (\bibinfo{year}{1971}).

\bibitem[{\citenamefont{McWhan et~al.}(1973)\citenamefont{McWhan, Menth,
  Remeika, Rice, and Brinkman}}]{mcw73}
\bibinfo{author}{\bibfnamefont{D.~B.} \bibnamefont{McWhan}},
  \bibinfo{author}{\bibfnamefont{A.}~\bibnamefont{Menth}},
  \bibinfo{author}{\bibfnamefont{J.~P.} \bibnamefont{Remeika}},
  \bibinfo{author}{\bibfnamefont{T.~M.} \bibnamefont{Rice}}, \bibnamefont{and}
  \bibinfo{author}{\bibfnamefont{W.~F.} \bibnamefont{Brinkman}},
  \bibinfo{journal}{Phys. Rev. B} \textbf{\bibinfo{volume}{7}},
  \bibinfo{pages}{1920} (\bibinfo{year}{1973}).

\bibitem[{\citenamefont{Fo{\"e}x}(1946)}]{foe46}
\bibinfo{author}{\bibfnamefont{M.}~\bibnamefont{Fo{\"e}x}},
  \bibinfo{journal}{Comptes rendus hebdomadaires des s\'eances de l'Acad\'emie
  des sciences} \textbf{\bibinfo{volume}{223}}, \bibinfo{pages}{1126}
  (\bibinfo{year}{1946}).

\bibitem[{\citenamefont{Moon}(1970)}]{moo70}
\bibinfo{author}{\bibfnamefont{R.~M.} \bibnamefont{Moon}},
  \bibinfo{journal}{Phys. Rev. Lett.} \textbf{\bibinfo{volume}{25}},
  \bibinfo{pages}{527} (\bibinfo{year}{1970}).

\bibitem[{\citenamefont{McWhan and Remeika}(1970)}]{mcw70}
\bibinfo{author}{\bibfnamefont{D.~B.} \bibnamefont{McWhan}} \bibnamefont{and}
  \bibinfo{author}{\bibfnamefont{J.~P.} \bibnamefont{Remeika}},
  \bibinfo{journal}{Phys. Rev. B} \textbf{\bibinfo{volume}{2}},
  \bibinfo{pages}{3734} (\bibinfo{year}{1970}).

\bibitem[{\citenamefont{Dernier and Marezio}(1970)}]{der70_2}
\bibinfo{author}{\bibfnamefont{P.~D.} \bibnamefont{Dernier}} \bibnamefont{and}
  \bibinfo{author}{\bibfnamefont{M.}~\bibnamefont{Marezio}},
  \bibinfo{journal}{Phys. Rev. B} \textbf{\bibinfo{volume}{2}},
  \bibinfo{pages}{3771} (\bibinfo{year}{1970}).

\bibitem[{\citenamefont{Grieger et~al.}(2012)\citenamefont{Grieger, Piefke,
  Peil, and Lechermann}}]{gri12}
\bibinfo{author}{\bibfnamefont{D.}~\bibnamefont{Grieger}},
  \bibinfo{author}{\bibfnamefont{C.}~\bibnamefont{Piefke}},
  \bibinfo{author}{\bibfnamefont{O.~E.} \bibnamefont{Peil}}, \bibnamefont{and}
  \bibinfo{author}{\bibfnamefont{F.}~\bibnamefont{Lechermann}},
  \bibinfo{journal}{Phys. Rev. B} \textbf{\bibinfo{volume}{86}},
  \bibinfo{pages}{155121} (\bibinfo{year}{2012}).

\bibitem[{\citenamefont{Ezhov et~al.}(1999)\citenamefont{Ezhov, Anisimov,
  Khomskii, and Sawatzky}}]{ezh99}
\bibinfo{author}{\bibfnamefont{S.~Y.} \bibnamefont{Ezhov}},
  \bibinfo{author}{\bibfnamefont{V.~I.} \bibnamefont{Anisimov}},
  \bibinfo{author}{\bibfnamefont{D.~I.} \bibnamefont{Khomskii}},
  \bibnamefont{and} \bibinfo{author}{\bibfnamefont{G.~A.}
  \bibnamefont{Sawatzky}}, \bibinfo{journal}{Phys. Rev. Lett.}
  \textbf{\bibinfo{volume}{83}}, \bibinfo{pages}{4136} (\bibinfo{year}{1999}).

\bibitem[{\citenamefont{Tenailleau et~al.}(2003)\citenamefont{Tenailleau,
  Suard, Rodriguez-Carvajal, Gibaud, and Lacorre}}]{ten03}
\bibinfo{author}{\bibfnamefont{C.}~\bibnamefont{Tenailleau}},
  \bibinfo{author}{\bibfnamefont{E.}~\bibnamefont{Suard}},
  \bibinfo{author}{\bibfnamefont{J.}~\bibnamefont{Rodriguez-Carvajal}},
  \bibinfo{author}{\bibfnamefont{A.}~\bibnamefont{Gibaud}}, \bibnamefont{and}
  \bibinfo{author}{\bibfnamefont{P.}~\bibnamefont{Lacorre}},
  \bibinfo{journal}{Journal of Solid State Chemistry}
  \textbf{\bibinfo{volume}{174}}, \bibinfo{pages}{431} (\bibinfo{year}{2003}).

\bibitem[{\citenamefont{Rozier et~al.}(2002)\citenamefont{Rozier, Ratuszna, and
  Galy}}]{roz02}
\bibinfo{author}{\bibfnamefont{P.}~\bibnamefont{Rozier}},
  \bibinfo{author}{\bibfnamefont{A.}~\bibnamefont{Ratuszna}}, \bibnamefont{and}
  \bibinfo{author}{\bibfnamefont{J.}~\bibnamefont{Galy}},
  \bibinfo{journal}{Zeitschrift f\"ur anorganische und allgemeine Chemie}
  \textbf{\bibinfo{volume}{628}}, \bibinfo{pages}{1236} (\bibinfo{year}{2002}).

\bibitem[{\citenamefont{Castellani
  et~al.}(1978{\natexlab{a}})\citenamefont{Castellani, Natoli, and
  Ranninger}}]{cas78_2}
\bibinfo{author}{\bibfnamefont{C.}~\bibnamefont{Castellani}},
  \bibinfo{author}{\bibfnamefont{C.~R.} \bibnamefont{Natoli}},
  \bibnamefont{and}
  \bibinfo{author}{\bibfnamefont{J.}~\bibnamefont{Ranninger}},
  \bibinfo{journal}{Phys. Rev. B} \textbf{\bibinfo{volume}{18}},
  \bibinfo{pages}{4967} (\bibinfo{year}{1978}{\natexlab{a}}).

\bibitem[{\citenamefont{Castellani
  et~al.}(1978{\natexlab{b}})\citenamefont{Castellani, Natoli, and
  Ranninger}}]{cas78}
\bibinfo{author}{\bibfnamefont{C.}~\bibnamefont{Castellani}},
  \bibinfo{author}{\bibfnamefont{C.~R.} \bibnamefont{Natoli}},
  \bibnamefont{and}
  \bibinfo{author}{\bibfnamefont{J.}~\bibnamefont{Ranninger}},
  \bibinfo{journal}{Phys. Rev. B} \textbf{\bibinfo{volume}{18}},
  \bibinfo{pages}{4945} (\bibinfo{year}{1978}{\natexlab{b}}).

\bibitem[{\citenamefont{Castellani
  et~al.}(1978{\natexlab{c}})\citenamefont{Castellani, Natoli, and
  Ranninger}}]{cas78_3}
\bibinfo{author}{\bibfnamefont{C.}~\bibnamefont{Castellani}},
  \bibinfo{author}{\bibfnamefont{C.~R.} \bibnamefont{Natoli}},
  \bibnamefont{and}
  \bibinfo{author}{\bibfnamefont{J.}~\bibnamefont{Ranninger}},
  \bibinfo{journal}{Phys. Rev. B} \textbf{\bibinfo{volume}{18}},
  \bibinfo{pages}{5001} (\bibinfo{year}{1978}{\natexlab{c}}).

\bibitem[{\citenamefont{Kugel' and Khomskii}(1982)}]{kug82}
\bibinfo{author}{\bibfnamefont{K.~I.} \bibnamefont{Kugel'}} \bibnamefont{and}
  \bibinfo{author}{\bibfnamefont{D.~I.} \bibnamefont{Khomskii}},
  \bibinfo{journal}{Sov.~Phys.~Usp.} \textbf{\bibinfo{volume}{25}},
  \bibinfo{pages}{231} (\bibinfo{year}{1982}).

\bibitem[{\citenamefont{Park et~al.}(2000)\citenamefont{Park, Tjeng, Tanaka,
  Allen, Chen, Metcalf, Honig, de~Groot, and Sawatzky}}]{par00}
\bibinfo{author}{\bibfnamefont{J.-H.} \bibnamefont{Park}},
  \bibinfo{author}{\bibfnamefont{L.~H.} \bibnamefont{Tjeng}},
  \bibinfo{author}{\bibfnamefont{A.}~\bibnamefont{Tanaka}},
  \bibinfo{author}{\bibfnamefont{J.~W.} \bibnamefont{Allen}},
  \bibinfo{author}{\bibfnamefont{C.~T.} \bibnamefont{Chen}},
  \bibinfo{author}{\bibfnamefont{P.}~\bibnamefont{Metcalf}},
  \bibinfo{author}{\bibfnamefont{J.~M.} \bibnamefont{Honig}},
  \bibinfo{author}{\bibfnamefont{F.~M.~F.} \bibnamefont{de~Groot}},
  \bibnamefont{and} \bibinfo{author}{\bibfnamefont{G.~A.}
  \bibnamefont{Sawatzky}}, \bibinfo{journal}{Phys. Rev. B}
  \textbf{\bibinfo{volume}{61}}, \bibinfo{pages}{11506} (\bibinfo{year}{2000}).

\bibitem[{\citenamefont{Paolasini et~al.}(1999)\citenamefont{Paolasini,
  Vettier, de~Bergevin, Yakhou, Mannix, Stunault, Neubeck, Altarelli, Fabrizio,
  Metcalf et~al.}}]{pao99}
\bibinfo{author}{\bibfnamefont{L.}~\bibnamefont{Paolasini}},
  \bibinfo{author}{\bibfnamefont{C.}~\bibnamefont{Vettier}},
  \bibinfo{author}{\bibfnamefont{F.}~\bibnamefont{de~Bergevin}},
  \bibinfo{author}{\bibfnamefont{F.}~\bibnamefont{Yakhou}},
  \bibinfo{author}{\bibfnamefont{D.}~\bibnamefont{Mannix}},
  \bibinfo{author}{\bibfnamefont{A.}~\bibnamefont{Stunault}},
  \bibinfo{author}{\bibfnamefont{W.}~\bibnamefont{Neubeck}},
  \bibinfo{author}{\bibfnamefont{M.}~\bibnamefont{Altarelli}},
  \bibinfo{author}{\bibfnamefont{M.}~\bibnamefont{Fabrizio}},
  \bibinfo{author}{\bibfnamefont{P.~A.} \bibnamefont{Metcalf}},
  \bibnamefont{et~al.}, \bibinfo{journal}{Phys. Rev. Lett.}
  \textbf{\bibinfo{volume}{82}}, \bibinfo{pages}{4719} (\bibinfo{year}{1999}).

\bibitem[{\citenamefont{Mila et~al.}(2000)\citenamefont{Mila, Shiina, Zhang,
  Joshi, Ma, Anisimov, and Rice}}]{mil00}
\bibinfo{author}{\bibfnamefont{F.}~\bibnamefont{Mila}},
  \bibinfo{author}{\bibfnamefont{R.}~\bibnamefont{Shiina}},
  \bibinfo{author}{\bibfnamefont{F.-C.} \bibnamefont{Zhang}},
  \bibinfo{author}{\bibfnamefont{A.}~\bibnamefont{Joshi}},
  \bibinfo{author}{\bibfnamefont{M.}~\bibnamefont{Ma}},
  \bibinfo{author}{\bibfnamefont{V.}~\bibnamefont{Anisimov}}, \bibnamefont{and}
  \bibinfo{author}{\bibfnamefont{T.~M.} \bibnamefont{Rice}},
  \bibinfo{journal}{Phys. Rev. Lett.} \textbf{\bibinfo{volume}{85}},
  \bibinfo{pages}{1714} (\bibinfo{year}{2000}).

\bibitem[{\citenamefont{Di~Matteo et~al.}(2002)\citenamefont{Di~Matteo,
  Perkins, and Natoli}}]{mat02}
\bibinfo{author}{\bibfnamefont{S.}~\bibnamefont{Di~Matteo}},
  \bibinfo{author}{\bibfnamefont{N.~B.} \bibnamefont{Perkins}},
  \bibnamefont{and} \bibinfo{author}{\bibfnamefont{C.~R.}
  \bibnamefont{Natoli}}, \bibinfo{journal}{Journal of Physics: Condensed
  Matter} \textbf{\bibinfo{volume}{14}}, \bibinfo{pages}{L37}
  (\bibinfo{year}{2002}).

\bibitem[{\citenamefont{Tanaka}(2002)}]{tan02}
\bibinfo{author}{\bibfnamefont{A.}~\bibnamefont{Tanaka}},
  \bibinfo{journal}{Journal of the Physical Society of Japan}
  \textbf{\bibinfo{volume}{71}}, \bibinfo{pages}{1091} (\bibinfo{year}{2002}).

\bibitem[{\citenamefont{Saha-Dasgupta et~al.}(2009)\citenamefont{Saha-Dasgupta,
  Andersen, Nuss, Poteryaev, Georges, and Lichtenstein}}]{sah09}
\bibinfo{author}{\bibfnamefont{T.}~\bibnamefont{Saha-Dasgupta}},
  \bibinfo{author}{\bibfnamefont{O.~K.} \bibnamefont{Andersen}},
  \bibinfo{author}{\bibfnamefont{J.}~\bibnamefont{Nuss}},
  \bibinfo{author}{\bibfnamefont{A.~I.} \bibnamefont{Poteryaev}},
  \bibinfo{author}{\bibfnamefont{A.}~\bibnamefont{Georges}}, \bibnamefont{and}
  \bibinfo{author}{\bibfnamefont{A.~I.} \bibnamefont{Lichtenstein}},
  \bibinfo{journal}{arXiv:0907.2841}  (\bibinfo{year}{2009}).

\bibitem[{\citenamefont{Elfimov et~al.}(2003)\citenamefont{Elfimov,
  Saha-Dasgupta, and Korotin}}]{elf03}
\bibinfo{author}{\bibfnamefont{I.~S.} \bibnamefont{Elfimov}},
  \bibinfo{author}{\bibfnamefont{T.}~\bibnamefont{Saha-Dasgupta}},
  \bibnamefont{and} \bibinfo{author}{\bibfnamefont{M.~A.}
  \bibnamefont{Korotin}}, \bibinfo{journal}{Phys. Rev. B}
  \textbf{\bibinfo{volume}{68}}, \bibinfo{pages}{113105}
  (\bibinfo{year}{2003}).

\bibitem[{\citenamefont{Perkins et~al.}(2009)\citenamefont{Perkins, Di~Matteo,
  and Natoli}}]{mat09}
\bibinfo{author}{\bibfnamefont{N.~B.} \bibnamefont{Perkins}},
  \bibinfo{author}{\bibfnamefont{S.}~\bibnamefont{Di~Matteo}},
  \bibnamefont{and} \bibinfo{author}{\bibfnamefont{C.~R.}
  \bibnamefont{Natoli}}, \bibinfo{journal}{Phys. Rev. B}
  \textbf{\bibinfo{volume}{80}}, \bibinfo{pages}{165106}
  (\bibinfo{year}{2009}).

\bibitem[{\citenamefont{Di~Matteo}(2005)}]{mat05}
\bibinfo{author}{\bibfnamefont{S.}~\bibnamefont{Di~Matteo}},
  \bibinfo{journal}{Physica Scripta} \textbf{\bibinfo{volume}{71}},
  \bibinfo{pages}{CC1} (\bibinfo{year}{2005}).

\bibitem[{\citenamefont{Held et~al.}(2001)\citenamefont{Held, Keller, Eyert,
  Vollhardt, and Anisimov}}]{hel01}
\bibinfo{author}{\bibfnamefont{K.}~\bibnamefont{Held}},
  \bibinfo{author}{\bibfnamefont{G.}~\bibnamefont{Keller}},
  \bibinfo{author}{\bibfnamefont{V.}~\bibnamefont{Eyert}},
  \bibinfo{author}{\bibfnamefont{D.}~\bibnamefont{Vollhardt}},
  \bibnamefont{and} \bibinfo{author}{\bibfnamefont{V.~I.}
  \bibnamefont{Anisimov}}, \bibinfo{journal}{Phys. Rev. Lett.}
  \textbf{\bibinfo{volume}{86}}, \bibinfo{pages}{5345} (\bibinfo{year}{2001}).

\bibitem[{\citenamefont{Perdew et~al.}(1996)\citenamefont{Perdew, Burke, and
  Ernzerhof}}]{per96}
\bibinfo{author}{\bibfnamefont{J.~P.} \bibnamefont{Perdew}},
  \bibinfo{author}{\bibfnamefont{K.}~\bibnamefont{Burke}}, \bibnamefont{and}
  \bibinfo{author}{\bibfnamefont{M.}~\bibnamefont{Ernzerhof}},
  \bibinfo{journal}{Phys. Rev. Lett.} \textbf{\bibinfo{volume}{77}},
  \bibinfo{pages}{3865} (\bibinfo{year}{1996}).

\bibitem[{\citenamefont{Giannozzi et~al.}(2009)\citenamefont{Giannozzi, Baroni,
  Bonini, Calandra, Car, Cavazzoni, Ceresoli, Chiarotti, Cococcioni, Dabo
  et~al.}}]{gia09}
\bibinfo{author}{\bibfnamefont{P.}~\bibnamefont{Giannozzi}},
  \bibinfo{author}{\bibfnamefont{S.}~\bibnamefont{Baroni}},
  \bibinfo{author}{\bibfnamefont{N.}~\bibnamefont{Bonini}},
  \bibinfo{author}{\bibfnamefont{M.}~\bibnamefont{Calandra}},
  \bibinfo{author}{\bibfnamefont{R.}~\bibnamefont{Car}},
  \bibinfo{author}{\bibfnamefont{C.}~\bibnamefont{Cavazzoni}},
  \bibinfo{author}{\bibfnamefont{D.}~\bibnamefont{Ceresoli}},
  \bibinfo{author}{\bibfnamefont{G.~L.} \bibnamefont{Chiarotti}},
  \bibinfo{author}{\bibfnamefont{M.}~\bibnamefont{Cococcioni}},
  \bibinfo{author}{\bibfnamefont{I.}~\bibnamefont{Dabo}}, \bibnamefont{et~al.},
  \bibinfo{journal}{Journal of Physics: Condensed Matter}
  \textbf{\bibinfo{volume}{21}}, \bibinfo{pages}{395502}
  (\bibinfo{year}{2009}), \urlprefix\url{http://www.quantum-espresso.org}.

\bibitem[{\citenamefont{Dernier}(1970)}]{der70}
\bibinfo{author}{\bibfnamefont{P.~D.} \bibnamefont{Dernier}},
  \bibinfo{journal}{J. Phys. Chem. Solids} \textbf{\bibinfo{volume}{31}},
  \bibinfo{pages}{2569} (\bibinfo{year}{1970}).

\bibitem[{\citenamefont{Ding et~al.}(2014)\citenamefont{Ding, Chen, Zeng, Kim,
  Han, Balasubramanian, Gordon, Li, Bai, Popov et~al.}}]{din14}
\bibinfo{author}{\bibfnamefont{Y.}~\bibnamefont{Ding}},
  \bibinfo{author}{\bibfnamefont{C.-C.} \bibnamefont{Chen}},
  \bibinfo{author}{\bibfnamefont{Q.}~\bibnamefont{Zeng}},
  \bibinfo{author}{\bibfnamefont{H.-S.} \bibnamefont{Kim}},
  \bibinfo{author}{\bibfnamefont{M.~J.} \bibnamefont{Han}},
  \bibinfo{author}{\bibfnamefont{M.}~\bibnamefont{Balasubramanian}},
  \bibinfo{author}{\bibfnamefont{R.}~\bibnamefont{Gordon}},
  \bibinfo{author}{\bibfnamefont{F.}~\bibnamefont{Li}},
  \bibinfo{author}{\bibfnamefont{L.}~\bibnamefont{Bai}},
  \bibinfo{author}{\bibfnamefont{D.}~\bibnamefont{Popov}},
  \bibnamefont{et~al.}, \bibinfo{journal}{Phys. Rev. Lett.}
  \textbf{\bibinfo{volume}{112}}, \bibinfo{pages}{056401}
  (\bibinfo{year}{2014}).

\bibitem[{\citenamefont{Mattheiss}(1994)}]{mat94}
\bibinfo{author}{\bibfnamefont{L.~F.} \bibnamefont{Mattheiss}},
  \bibinfo{journal}{J. Phys.: Condens. Matter} \textbf{\bibinfo{volume}{6}},
  \bibinfo{pages}{6477} (\bibinfo{year}{1994}).

\bibitem[{\citenamefont{Eyert et~al.}(2005)\citenamefont{Eyert,
  Schwingenschl\"ogl, and Eckern}}]{eye05}
\bibinfo{author}{\bibfnamefont{V.}~\bibnamefont{Eyert}},
  \bibinfo{author}{\bibfnamefont{U.}~\bibnamefont{Schwingenschl\"ogl}},
  \bibnamefont{and} \bibinfo{author}{\bibfnamefont{U.}~\bibnamefont{Eckern}},
  \bibinfo{journal}{EPL (Europhysics Letters)} \textbf{\bibinfo{volume}{70}},
  \bibinfo{pages}{782} (\bibinfo{year}{2005}).

\bibitem[{\citenamefont{Grieger and Lechermann}(2014)}]{gri14}
\bibinfo{author}{\bibfnamefont{D.}~\bibnamefont{Grieger}} \bibnamefont{and}
  \bibinfo{author}{\bibfnamefont{F.}~\bibnamefont{Lechermann}},
  \bibinfo{journal}{Phys. Rev. B} \textbf{\bibinfo{volume}{90}},
  \bibinfo{pages}{115115} (\bibinfo{year}{2014}).

\bibitem[{\citenamefont{Cococcioni and de~Gironcoli}(2005)}]{coc05}
\bibinfo{author}{\bibfnamefont{M.}~\bibnamefont{Cococcioni}} \bibnamefont{and}
  \bibinfo{author}{\bibfnamefont{S.}~\bibnamefont{de~Gironcoli}},
  \bibinfo{journal}{Phys. Rev. B} \textbf{\bibinfo{volume}{71}},
  \bibinfo{pages}{035105} (\bibinfo{year}{2005}).

\bibitem[{\citenamefont{Abragam and Bleaney}(1970)}]{abr70}
\bibinfo{author}{\bibfnamefont{A.}~\bibnamefont{Abragam}} \bibnamefont{and}
  \bibinfo{author}{\bibfnamefont{B.}~\bibnamefont{Bleaney}},
  \emph{\bibinfo{title}{Introduction to ligand field theory}}
  (\bibinfo{publisher}{Clarendon Press - Oxford}, \bibinfo{year}{1970}), pp.
  \bibinfo{pages}{377--378 and 426--429}.

\bibitem[{\citenamefont{Cole and Garrett}(1970)}]{col70}
\bibinfo{author}{\bibfnamefont{G.~M.} \bibnamefont{Cole}} \bibnamefont{and}
  \bibinfo{author}{\bibfnamefont{B.~B.} \bibnamefont{Garrett}},
  \bibinfo{journal}{Inorganic Chemistry} \textbf{\bibinfo{volume}{9}},
  \bibinfo{pages}{1898} (\bibinfo{year}{1970}).

\bibitem[{\citenamefont{Tchernyshyov}(2004)}]{tch04}
\bibinfo{author}{\bibfnamefont{O.}~\bibnamefont{Tchernyshyov}},
  \bibinfo{journal}{Phys. Rev. Lett.} \textbf{\bibinfo{volume}{93}},
  \bibinfo{pages}{157206} (\bibinfo{year}{2004}).

\bibitem[{\citenamefont{Perkins and Sikora}(2007)}]{per07}
\bibinfo{author}{\bibfnamefont{N.~B.} \bibnamefont{Perkins}} \bibnamefont{and}
  \bibinfo{author}{\bibfnamefont{O.}~\bibnamefont{Sikora}},
  \bibinfo{journal}{Phys. Rev. B} \textbf{\bibinfo{volume}{76}},
  \bibinfo{pages}{214434} (\bibinfo{year}{2007}).

\bibitem[{\citenamefont{Dal~Corso and Mosca~Conte}(2005)}]{cor05}
\bibinfo{author}{\bibfnamefont{A.}~\bibnamefont{Dal~Corso}} \bibnamefont{and}
  \bibinfo{author}{\bibfnamefont{A.}~\bibnamefont{Mosca~Conte}},
  \bibinfo{journal}{Phys. Rev. B} \textbf{\bibinfo{volume}{71}},
  \bibinfo{pages}{115106} (\bibinfo{year}{2005}).

\bibitem[{\citenamefont{Liechtenstein et~al.}(1995)\citenamefont{Liechtenstein,
  Anisimov, and Zaanen}}]{lie95}
\bibinfo{author}{\bibfnamefont{A.~I.} \bibnamefont{Liechtenstein}},
  \bibinfo{author}{\bibfnamefont{V.~I.} \bibnamefont{Anisimov}},
  \bibnamefont{and} \bibinfo{author}{\bibfnamefont{J.}~\bibnamefont{Zaanen}},
  \bibinfo{journal}{Phys. Rev. B} \textbf{\bibinfo{volume}{52}},
  \bibinfo{pages}{R5467} (\bibinfo{year}{1995}).

\bibitem[{\citenamefont{Albuquerque et~al.}(2011)\citenamefont{Albuquerque,
  Schwandt, Het\'enyi, Capponi, Mambrini, and L\"auchli}}]{alb11}
\bibinfo{author}{\bibfnamefont{A.}~\bibnamefont{Albuquerque}},
  \bibinfo{author}{\bibfnamefont{D.}~\bibnamefont{Schwandt}},
  \bibinfo{author}{\bibfnamefont{B.}~\bibnamefont{Het\'enyi}},
  \bibinfo{author}{\bibfnamefont{S.}~\bibnamefont{Capponi}},
  \bibinfo{author}{\bibfnamefont{M.}~\bibnamefont{Mambrini}}, \bibnamefont{and}
  \bibinfo{author}{\bibfnamefont{A.}~\bibnamefont{L\"auchli}},
  \bibinfo{journal}{Phys. Rev. B} \textbf{\bibinfo{volume}{84}},
  \bibinfo{pages}{024406} (\bibinfo{year}{2011}).

\bibitem[{\citenamefont{Fouet et~al.}(2001)\citenamefont{Fouet, Sindzingre, and
  Lhuillier}}]{fou01}
\bibinfo{author}{\bibfnamefont{J.}~\bibnamefont{Fouet}},
  \bibinfo{author}{\bibfnamefont{P.}~\bibnamefont{Sindzingre}},
  \bibnamefont{and}
  \bibinfo{author}{\bibfnamefont{C.}~\bibnamefont{Lhuillier}},
  \bibinfo{journal}{The European Physical Journal B - Condensed Matter and
  Complex Systems} \textbf{\bibinfo{volume}{20}}, \bibinfo{pages}{241}
  (\bibinfo{year}{2001}).

\bibitem[{\citenamefont{Mezzacapo and Boninsegni}(2012)}]{mez12}
\bibinfo{author}{\bibfnamefont{F.}~\bibnamefont{Mezzacapo}} \bibnamefont{and}
  \bibinfo{author}{\bibfnamefont{M.}~\bibnamefont{Boninsegni}},
  \bibinfo{journal}{Phys. Rev. B} \textbf{\bibinfo{volume}{85}},
  \bibinfo{pages}{060402} (\bibinfo{year}{2012}).

\end{thebibliography}

\end{document}